\documentclass[11pt]{article}
\usepackage{latexsym}
\usepackage{amssymb}

\def\qed{\hfill$\Box$}

\newcommand{\Xp}{\mbox{\boldmath $X$}}
\newcommand{\Cp}{\mbox{\boldmath $C$}}

\newcommand{\Zp}{\mbox{\boldmath $Z$}}

\newcommand{\bydef}{\mbox{$ \;\stackrel{\triangle}{=}\; $}}

\newcommand{\leqa}{\mbox{$ \;\stackrel{(a)}{\leq}\; $}}
\newcommand{\leqb}{\mbox{$ \;\stackrel{(b)}{\leq}\; $}}

\newcommand{\geqa}{\mbox{$ \;\stackrel{(a)}{\geq}\; $}}

\newcommand{\eqa}{\mbox{$ \;\stackrel{(a)}{=}\; $}}
\newcommand{\eqb}{\mbox{$ \;\stackrel{(b)}{=}\; $}}
\newcommand{\eqc}{\mbox{$ \;\stackrel{(c)}{=}\; $}}

\newcommand{\RL}{{\mathbb R}}

\newcommand{\IND}{{\mathbb I}}
\newcommand{\BBP}{{\mathbb P}}

\newcommand{\Qtilde}{\widetilde{Q}}
\newcommand{\PR}{\mbox{\rm Pr}}
\newcommand{\VAR}{\mbox{\rm Var}}

\newcommand{\essinf}{\mathop{\rm ess\, inf}}

\newcommand{\Ahat}{\hat{A}}
\newcommand{\Ahatn}{\mbox{$\hat{A}^n$}}

\newcommand{\Phatn}{\mbox{$\hat{P}_n$}}
\newcommand{\Phatyk}{\hat{P}_{Y_1^n(k)}}
\newcommand{\Qhatn}{\widehat{Q}_n}
\newcommand{\Qhat}{\widehat{Q}}

\newcommand{\calL}{\mbox{${\cal L}$}}

\newcommand{\Dmin}{\mbox{$D_{\rm min}$}}

\newcommand{\Qmax}{\mbox{$Q_{\rm max}$}}
\newcommand{\hmax}{\mbox{$h_{\rm max}$}}
\newcommand{\ymax}{\mbox{$y_{\rm max}$}}

\newcommand{\Dav}{\mbox{$D_{\rm av}$}}

\newcommand{\Wadd}{\mbox{$W_{\rm add}$}}
\newcommand{\LA}{\Lambda}
\newcommand{\THE}{\Theta}

\newcommand{\la}{\lambda}

\newcommand{\smalloneovern}{{\textstyle\frac{1}{n}}}
\newcommand{\expd}{E_n^{(\Delta)}}
\newcommand{\Gts}{\mbox{Geom}^*}

\newcommand{\iid}{\mbox{i.i.d.}\!}

\newcommand{\argmin}{\mathop{\rm arg\, min}}
\newcommand{\argmax}{\mathop{\rm arg\, max}}

\def\be{\begin{eqnarray}}
\def\ee{\end{eqnarray}}
\def\ben{\begin{eqnarray*}}
\def\een{\end{eqnarray*}}

\newlength{\noteWidth}
\long\def\notes#1{\ifinner
             {\tiny #1}
             \else
             \marginpar{\parbox[t]{\noteWidth}{\raggedright\tiny #1}}
             \fi}

\long\def\answer#1{\ifinner
             {\tiny #1}
             \else
             \marginpar{\parbox[t]{\noteWidth}{\raggedright\tiny\it #1}}
             \fi}

\title{Mismatched Codebooks and the Role of Entropy-Coding\\
in Lossy Data Compression}

\author{I.\ Kontoyiannis \and R.\ Zamir}

\date{\today}

\begin{document}
\bibliographystyle{plain}
\maketitle

\thispagestyle{empty}

\footnotetext[1]{
Ioannis Kontoyiannis is with the Division
of Applied Mathematics and the Department
of Computer Science, Brown University,
Box F, 182 George Street, Providence, RI 02912, USA.
Email: {\tt yiannis@dam.brown.edu}
Web: {\tt www.dam.brown.edu/people/yiannis/}
}

\footnotetext[2]{ Ram Zamir is with the Department
of Electrical Engineering - Systems, Tel-Aviv University,
Ramat Aviv 69978, Israel.
Email: {\tt zamir@eng.tau.ac.il} Web: {\tt www.eng.tau.ac.il/\~{}zamir/} }
\footnotetext[3]{
I.\ Kontoyiannis was supported in part
by NSF grant \#0073378-CCR,
        and by USDA-IFAFS grant
        \#00-52100-9615.
R.\ Zamir was supported in part by the US-Israel Bi-National
Science Foundation grant 1998-309.}


{\bf Abstract --- } We introduce a universal quantization scheme
based on random coding, and we analyze its performance.  This scheme
consists of a source-independent random codebook (typically {\em
mismatched} to the source distribution), followed by optimal
entropy-coding that is {\em matched} to the quantized codeword 
distribution. A single-letter formula is derived for the
rate achieved by this scheme at a given distortion, in the limit
of large codebook dimension.  The rate reduction due to
entropy-coding is quantified, and it is shown that it can be
arbitrarily large.
In the special case of ``almost uniform''
codebooks (e.g., an $\iid$ Gaussian codebook with large variance)
and difference distortion measures, a novel connection is drawn
between the compression achieved by the present scheme
and the performance of ``universal''
entropy-coded dithered lattice quantizers.
This connection generalizes the ``half-a-bit'' bound on the redundancy of
dithered lattice quantizers.
Moreover, it demonstrates a strong notion of universality
where a single ``almost uniform'' codebook is near-optimal
for {\em any} source and {\em any} difference distortion measure.
%
%
%
%
The proofs are based on the fact that the limiting
empirical distribution of the first matching
codeword in a random codebook can be precisely identified.
This is done using elaborate large deviations techniques,
that allow the derivation of a new ``almost sure''
version of the conditional limit theorem.

\medskip



\newpage

\section{Introduction}

\subsection{Mismatched Quantization and Compression}

Variable-rate lossless compression -- or {\em entropy-coding} --
is an efficient method for enhancing the compression
performance of quantizers \cite{gish-pierce:68,chou-l-gray}.
This paper investigates the role of entropy-coding when
the quantizer codebook is {\em mismatched} with respect
to the source distribution.  Our motivation mainly comes
from Ziv's concept of {\em universal quantization}
for lossy compression of real-valued sources with unknown
statistics \cite{ziv:1}. Ziv's scheme uses a randomized (``dithered'')
lattice quantizer, which is scaled to meet the target
distortion level, and the quantizer is followed by a universal
lossless encoder which reduces the coding rate to the
true entropy of the quantized sequence. Neuhoff
\cite{neuhoff:1} suggested that the universal quantizer
could be viewed as an efficient combination of a ``simple''
robust quantizer and a ``complex'' lossless encoder.
Variations on the problem of entropy-coded dithered
quantization (ECDQ) can be found
in \cite{gutman:87,feder-zamir:92,feder-zamir:96}.

Intuitively, the quantizer mismatch leaves much room for
rate savings using entropy-coding.  Moreover,
unlike {\em optimum} entropy-constrained vector
quantization (ECVQ) \cite{gersho:79,lookabaugh-gray:89},
the entropy-coding gain in the mismatched case does not vanish
even in the limit of large vector dimension.
For a rather trivial example, note that
the un-coded rate of an unbounded lattice quantizer is
infinite, but it becomes finite after entropy-coding
if the source has finite variance.
The advantage of entropy-coding a mismatched quantizer
is particularly prominent at high resolution
quantization conditions.
Gray and Linder show that {\em optimum} high-rate performance
for mean-squared distortion can be achieved even if the quantizer
codebook is mismatched with respect to the source
(specifically, if it is designed for a source with uniform
density), as long as the quantizer output is entropy-coded
according to the {\em true} quantizer output distribution
\cite[sec.~VII]{gray-linder:03}.
%
%
As we shall see here, similar behavior occurs at {\em any} resolution,
only with a slight rate loss due to the codebook mismatch.

One of the central results of universal quantization
theory is that, after entropy-coding, the rate loss of
the universal quantizer with respect to the optimum ECVQ
is bounded for {\em all} sources and
{\em all} distortion levels by a universal constant
\cite{ziv:1,feder-zamir:92}.
For example, for squared error distortion,
the rate loss of a $k$-dimensional lattice ECDQ is bounded by
$(1/2) \log(4 \pi e G_k)$ bits,
where $G_k$ is the normalized second moment of the lattice;
this bound is $\approx$ 0.754 bits for $k=1$,
and it converges to 1/2 bit as $k \rightarrow \infty$
(where $\log=\log_2$).
These results are limited, however, to lattice structured
quantizers, and more specifically to those lattice dimensions
and distortion measures which are covered by lattice coding theory.


The central goal of this paper is to develop a structure-free
framework for mismatched, entropy-coded quantization
at an arbitrary distortion level, based on random coding
ideas and techniques.
The random coding framework, although not constructive,
allows us to precisely quantify two important operational
quantities:
(a)~The potential rate gain due to entropy-coding when using
a mismatched random codebook; equivalently, this can be thought
of as the rate loss of the straightforward scheme which uses a
mismatched codebook without entropy-coding. (b)~The rate loss
due to quantizer mismatch, over the optimal rate-distortion
function: We will derive a universal upper bound for this
rate loss, 
analogous to the half-a-bit bound for lattice ECDQs
and quadratic distortion described above.

Mismatched random codebooks for {\em fixed-rate} lossy source
coding have been investigated by Sakrison
\cite{sakrison:69,sakrison:70},
Zhang and Wei \cite{zhang-wei:1},
Lapidoth \cite{lapidoth:97},
Zamir and Rose \cite{zamir-rose:01},
and others.
See \cite{kanlis:phd,kanlis-narayan-rimoldi}
and the references therein.
Specifically, source coding with a mismatched random
codebook (or string matching with a mismatched database)
has been considered
by Steinberg and Gutman \cite{steinberg-gutman},
Yang and Kieffer \cite{yang-kieffer:1}, and
Dembo and Kontoyiannis \cite{dembo-kontoyiannis:wyner},
among others. These works (and the references therein)
develop an extensive theory of mismatched
random lossy coding in the limit of large codebook dimension,
with an emphasis on precisely characterizing the
asymptotic rate and the redundancy of these schemes.
Here we continue that investigation, but we introduce
the additional step of entropy-coding the index
of the codebook before transmitting it to the decoder.


Entropy coding the codeword index in a source-matched
random lossy codebook has been considered in the early
work by Pinkston \cite{pinkston:67}.
Mismatched high resolution quantization has been considered
by Bucklew \cite{bucklew:84} and by Gray and Linder
\cite{gray-linder:03}, where several results, some
of which parallel those derived here, are presented.
Preliminary results on the entropy rates achieved by
mismatched random codebooks for general (non vanishing)
distortions and discrete memoryless
sources appear in \cite{zamir:index}.
Here we strengthen these results, and extend them
to richer classes of sources and codebook distributions.
In particular, we establish a formal connection between
ECDQ and entropy-coded random codebooks.

\subsection{Discussion of Main Results}

We begin in Section~2, where we derive asymptotic single-letter
characterizations for the compression rate
achieved by two different coding schemes,
both based on a random codebook
$\Cp_n = \{ Y_1^n(i), \ i=1,2,\ldots \}$
consisting of $\iid$ $n$-dimensional words $Y_1^n$,
each having $\iid$ components generated by
an arbitrary distribution $Q$.
We shall discuss later specific interesting choices
for the codebook distribution $Q$.
A natural motivation for the use of a mismatched $Q$
is the observation that, in many important applications,
the source statistics are generally unknown {\em a priori}
or they change with time -- or both.
%

Given a source string $X_1^n$ to be compressed
with distortion $D$ or less,
we consider the index $N_n$
of the first codeword in $\Cp_n$ that
matches $X_1^n$ within distortion $D$.
Our first result says that
as $n\to\infty$, the empirical
distribution of this first matching word
converges to a distribution $Q^*_{PQD}$
which can be identified as the solution of
an single-letter minimization problem.
[Here $P = P_1$ denotes the first-order marginal of
the source distribution.]
The proof is based on large deviations
techniques, and generalizes the ``favorite type
theorem'' of \cite{zamir-rose:01}.

Using this result we establish
an upper bound on the rate
achieved when such a random codebook
is used in conjunction with entropy-coding:
Suppose that the encoder first finds
the first $D$-close match at position
$N_n$, and then entropy-codes the
index $N_n$ conditional on the codebook
$\Cp_n$. The rate achieved for this
$D$-accurate description of $X_1^n$ is
$$H(N|\Cp)\bydef
\lim_{n\to\infty} \frac{1}{n} H(N_n |\Cp_n)
\;\;\;\;\mbox{bits/symbol}.$$
We then compare our bound with the limiting
rate $R(P,Q,D)$ achieved in the ``naive coding'' scenario,
where the encoder simply transmits the index
$N_n$ using Elias' code for the integers,
\[
R(P,Q,D)=\lim_{n\to\infty} \frac{1}{n} \log(N_n)
\;\;\;\;\mbox{bits/symbol}.
\]
We show that the rate gain of entropy-coding
over the naive coding scheme
(or, equivalently, the rate loss of naive
coding) satisfies
\[
\mbox{rate gain} =
R(P,Q,D) - H(N |\Cp) \geq H(Q^*_{PQD} \| Q)
\;\;\;\mbox{bits/symbol,}
\]
i.e., it is at least as large as the relative entropy
between the limiting empirical distribution of $Y_1^n(N_n)$
and the codebook-generating distribution $Q$.
For example, it is approximately $H(P \| Q)$
for small mean squared distortion $D$.
This lower bound is strictly positive, unless $Q$
is the optimal reproduction distribution
(i.e., the optimal output distribution of the
rate-distortion function).

This expression resembles the rate loss due to
mismatch in the {\em lossless} component of the code
at high resolution quantization
(see, e.g., \cite{gray-linder:03}).
Indeed, for small mean-squared distortion we have
$H(N |\Cp) \approx h(P) - \frac{1}{2} \log(2 \pi e D)$,
and
$R(P,Q,D) \approx h(P) - \frac{1}{2} \log(2 \pi e D) + H(P \| Q)$,
hence the latter amounts to encoding a source $\sim P$
using a code designed for a source $\sim Q$.
At {\em non-}high resolution conditions, however,
the rate loss remains positive even if the lossless
component of the code is matched, i.e.,
$H(N |\Cp)$ is in general strictly
above the rate-distortion function.

Of particular interest is the case of universal Gaussian
codebooks: Suppose we encode a real-valued
{\em memoryless} source using a white Gaussian codebook
$\sim N(0, \tau^2)$, with respect to squared error distortion.
If we simply use this codebook in the ``naive'' sense
described above, robust source
coding theory implies  that taking
$\tau^2 = \sigma^2 - D$,
where $\sigma^2$ denotes the source variance,
guarantees achieving the Gaussian rate-distortion
function
$R(D) = \frac{1}{2}\log(\sigma^2/D)$
for {\em any} source
\cite{sakrison:69,lapidoth:97}.
Since the Gaussian source is the hardest to compress in this class,
this implies high redundancy when the source is
far from Gaussian.

On the other hand, if we also allow the encoder
to entropy code the index,
the results are fundamentally different.
If we take the codebook variance $\tau^2$ to be large,
the codebook distribution becomes flat and it
is tempting to think that the codebook
itself looks approximately like the
codebook of a lattice quantizer. Indeed,
we show that as $\tau^2\to\infty$ the
rate $H(N|\Cp)$ achieved by
entropy-coding this Gaussian codebook
is no greater than the rate of a dithered
lattice quantizer with large lattice
dimension, given by
\ben
\limsup_{\tau^2 \rightarrow \infty}
H(N |\Cp) \leq I(X; X + Z_D)
\;\;\;\;\mbox{bits/symbol},
\een
where $I(X ;X + Z_D)$ denotes the mutual information
between the first source symbol $X$
and $X + Z_D$, where $Z_D$ is an independent
$N(0,D)$ random variable.
Combining this with well-known facts about
universal quantizers \cite{ziv:1,feder-zamir:92},
it follows that the naive coding rate is
going to infinity as $\tau^2 \rightarrow \infty$,
whereas the limiting rate achieved
by entropy-coding, $I(X; X + Z_D)$,
is at most $1/2$ bit above the rate-distortion
function of $X$,
and it coincides with the rate-distortion function
of $X$ in the limit of small $D$.
This new derivation provides an interesting
bridge between universal quantization theory
and mismatched random coding.

As observed by Yang and Kieffer \cite{yang-kieffer:1},
the naive coding rate $R(P,Q,D)$ depends only on the first-order
marginal of
the source distribution, hence it does not benefit from
memory in the source.
Moreover, as shown in detail in the sequel,
the entire dependence of the naive coding
rate on the source distribution $P$ may be very weak
(in fact, sometimes it is {\em entirely independent}
 of $P$), thus it is far from being optimal.
As we shall see, these disadvantages are
eliminated by the use of entropy coding.


In particular, for the case of
memoryless Gaussian codebooks with large variance,
we argue that the entropy-coding scheme achieves
a rate no greater than the mutual information rate
$I(\Xp,\Xp+\Zp_D)$, where $\Zp_D$ is an independent
white Gaussian process with variance $D$.
As before, this in turn implies that the
rate of the entropy-coded scheme is no greater
than $R(D)+1/2$ bits/symbol, where $R(D)$ is
the rate distortion function of the entire
source (not just the first-order rate-distortion
function).

We also show that these results generalize
beyond the Gaussian codebook case to a much wider class of codebooks,
namely, ``approximately flat'' codebooks with
distributions of exponential type, and to general
difference distortion measures.
We quantify the entropy-coding gain in this case, and
show that the resulting compression rate is
bounded above by $R(D)+C^*$ bits/symbol,
where $R(D)$ is the rate-distortion function
of the source, and $C^*$
is an upper bound for the ``min-max capacity''
defined in \cite{zamirWZ}. This is a new
generalization of the well-known
half-a-bit bound derived for dithered lattice
quantizers and squared distortion \cite{feder-zamir:92}
to the case of general difference distortion measures.
Moreover, {\em it implies the existence of a
single ensemble of codebooks which is
universal with respect to both the source
distribution and the distortion criterion},
resembling the results of Yang and Kieffer in
\cite{yang-kieffer:96}.

\medskip

The paper is organized as follows.
In Section~\ref{s:performance} we describe in detail
the entropy-coding scenario and the
naive coding scheme based on a mismatched
random codebook, and we state the
main result on the index entropy in Theorem~1.
Section~\ref{s:compare} contains two important
examples illustrating the entropy-coding gain,
including the case of universal Gaussian codebooks
mentioned above. In Section~\ref{gibbs} we state and prove
an almost sure conditional limit theorem (Theorem~3),
which forms the basis for the favorite type theorem
(Theorem~2) and for the proof of Theorem~1 which
is given in Section~\ref{proof}. Finally, in Section~\ref{section:last} we
give tighter bounds on the index entropy and
the entropy-coding gain for sources with memory.

\section{The Performance of Mismatched Codebooks}
\label{s:performance}

In this section we characterize the compression performance
achieved by memoryless random codebooks when used to
compress data generated from a stationary ergodic source.
Two coding scenarios are considered: The ``naive coding''
scenario where data is simply described by the index of
the first match in the codebook, and the ``entropy-coded''
case where this index is entropy-coded.

In Section~\ref{s:compare} we compare the performance
of these two schemes, and explicitly evaluate the entropy-coding
gain in two important special cases.

\subsection{Notation and Definitions}
We begin by introducing some basic definitions
and notation that will remain in effect for
the rest of the paper.

Consider a
stationary ergodic
process (or source)
$\Xp=\{X_n\;;\;n\geq 1\}$
taking values in the source alphabet $A$.
We will assume throughout that
$A$ is a complete, separable metric space
(often called a Polish space),
equipped with its associated Borel
$\sigma$-field ${\cal A}$. For the sake
of simplicity we also make the (rather
harmless) assumption that all singletons
are measurable, i.e.,
$\{x\}\in{\cal A}$ for all $x\in A$.
Similarly, for the reproduction
alphabet $\Ahat$ we take $(\Ahat,\hat{\cal A})$
to be the Borel measurable space
corresponding to a complete,
separable metric (or Polish) space $\Ahat$
and assume that $\{y\}\in\hat{\cal A}$
for all $y\in\Ahat$.
We write
$X_i^j$ for the vector of random
variables $X_i^j=(X_i,X_{i+1},\ldots,X_j)$,
and similarly
$x_i^j=(x_i,x_{i+1},\ldots,x_j)\in
A^{j-i+1}$
for a realization of these random variables,
$-\infty\leq i\leq j\leq \infty$.
We let $P_n$ denote the marginal
distribution of $X_1^n$ on $A^n$
($n\geq 1$), and write $\BBP$
for the distribution of the whole
process.
We use $P$ for the first-order marginal $P_1$.

Given an arbitrary nonnegative (measurable) function
$\rho:A\times\Ahat\to[0,\infty)$,
define a sequence of
single-letter (or ``additive'') distortion measures
$\rho_n:A^n\times\Ahatn\to[0,\infty)$ by
\ben
\rho_n(x_1^n,y_1^n)\bydef\frac{1}{n}\sum_{i=1}^n\rho(x_i,y_i)
\;\;\;\;x_1^n\in A^n,\;y_1^n\in\Ahatn.
\een
For a distortion level $D\geq 0$ and
a source string $x_1^n\in A^n$,
we write
$B(x_1^n,D)$ for
the distortion-ball of radius $D$ around $x_1^n$:
$$B(x_1^n,D)=\{y_1^n\in\Ahatn\;:\;\rho_n(x_1^n,y_1^n)\leq D\}.$$

Throughout the paper, $\log$ denotes
the logarithm to base 2 and $\ln$
denotes the natural logarithm. Unless otherwise
mentioned, all familiar information-theoretic
quantities (entropy, mutual information,
and so on)
are defined in terms
of logarithms taken to base 2,
and are therefore expressed in bits.

\subsection{Random Codebooks}
Given a probability measure $Q$ on the reproduction
alphabet $\Ahat$, a
{\em memoryless random codebook $\Cp_n$ with distribution $Q$}
is an infinite sequence of $\iid$ random vectors
$Y_1^n(i),$ $i\geq 1$,
with each
$Y_1^n(i)$
being distributed
according to the product measure
$Q^n$ on $\Ahatn$. In other words,
the components of
$Y_1^n(i)$ are $\iid$ with
distribution $Q$.
We write
$$\Cp_n\bydef\{Y_1^n(i)\;;\;i\geq 1\}$$
for the entire codebook,
and we call $Q$ the {\em codebook distribution.}

Suppose that, for a fixed $n$,
this codebook is available to both the encoder and decoder.
Given a distortion level $D$ and a source string $X_1^n$ to
be described with distortion $D$ or less,
the encoder looks for a $D$-close match of
$X_1^n$ into the codebook $\Cp_n$.
Let $N_n$ be the position of the first such match,
\ben
N_n\bydef \inf \{i\geq 1\;:\;\rho_n(X_1^n,Y_1^n(i))\leq D\},
\een
with the convention that the infimum of
the empty set equals $+\infty$.
Roughly speaking, the way the encoder describes
$X_1^n$
is by describing the position $N_n$ of this
first match.

Given a codebook distribution
$Q$ on $\Ahat$, we define
\ben
\Dmin & \bydef &  E_P[\essinf_{Y\sim Q} \;\rho(X,Y)]
    \\
\Dav & \bydef &  E_{P\times Q}[\rho(X,Y)],
\een
where $P=P_1$ denotes the first-order
marginal of $\Xp$.\footnote{Recall that
the essential infimum of
a function $g(Y)$ of the random variable
$Y$ with distribution $Q$ is defined as
$\essinf_{Y\sim Q} g(Y) =
\sup\{t\in\RL\;:\;Q\{g(Y)>t\}=1\}.$}
We will assume throughout that $\Dav$ is
finite.
Clearly $0\leq\Dmin\leq\Dav$.
To avoid the trivial case when $\rho(x,y)$
is constant for
($P$-almost) all
$x\in A$, we assume that with positive
$P$-probability $\rho(x,y)$ is not
essentially constant in $y$, that is:
\ben
\Dmin < \Dav.
\een
Note also that
for $D$ greater than $\Dav$
the rate-distortion function
$R(D)$ of $\Xp$ is zero, and
that for $D$ below $\Dmin$
no match can ever be found.
Therefore, from now on we
restrict our attention
to the interesting
range of distortion
levels $D\in(\Dmin,\Dav)$.

We consider two possible ways in which
the encoder can transmit $N_n$:
The simplest thing to do is
describe $N_n$ directly, using
some predetermined
code for the positive integers; see, e.g.,
\cite{elias}. This can be done with
approximately $\log( N_n)$ bits.
Alternatively, once the codebook
$\Cp_n$ has been fixed, the encoder
may choose to ``entropy-code''
$N_n$, giving it an average
description length of roughly $H(N_n|\Cp_n)$ bits.
This is equivalent to re-ordering the codewords according to
decreasing order of probabilities, and then describing the
new index $\pi(N_n)$ using approximately $\log (\pi(N_n))$ bits
like above.
When the statistics of the source are a-priori unknown,
we use a universal algorithm to entropy-code (or re-order)
the codewords.

\subsection{Naive Coding}
First we consider the case when the
encoder describes the index $N_n$
without entropy-coding; we refer to this
scenario as ``naive coding.''
As mentioned in the Introduction,
this coding scheme (and many variations
on it) has been analyzed extensively
in \cite{steinberg-gutman}\cite{yang-kieffer:1}%
\cite{dembo-kontoyiannis:wyner} and several other
works cited therein.
To avoid potentially infinite searches
in the codebook, we make the
simplifying assumption that the
encoder only describes $N_n$ when
it is smaller than $2^{nb}$,
where $b$ is some positive constant
to be chosen later. Accordingly,
we define the truncated index $N_n'$:
$$
N_n'
\bydef
\left
\{ \begin{array}{ll}
   N_n,
        & \mbox{if $N_n\leq \lfloor 2^{nb}\rfloor$}, \\
   \lfloor 2^{nb}\rfloor+1,    & \mbox{otherwise.}
\end{array}
\right.
$$
When $N_n$ exceeds $\lfloor 2^{nb}\rfloor$,
the encoder uses an alternative
description for $X_1^n$.
In order
to ensure that such a description
can be given with
finite rate, we introduce the
following simple conditions; cf.
\cite{kieffer:91,konto-zhang:02,dembo-kontoyiannis:wyner}.

\medskip

(WQC): For a distortion level $D\geq 0$ we say
that the {\em weak quantization condition (WQC)
holds at $D$} if there is a
(measurable) scalar quantizer $q:A\to B\subset\Ahat$
such that $B$ is a finite
or countably infinite set, and
$$\rho(x,q(x))\leq D\;\;\;\;\mbox{for all $x\in A.$}$$

($p$SQC): For a distortion level $D\geq 0$
we say that the
{\em $p$-strong quantization condition ($p$SQC)
holds at $D$} for some $p\geq 1$,
if (WQC) holds with respect to a
scalar quantizer $q$ also satisfying
\ben
M_p \; \bydef \;
\left\{E_P[(-\log \mu(q(X_1)))^p]\right\}^{1/p}<\infty,
\een
where $\mu$ denotes the (discrete)
distribution of the quantized random
variable $q(X)$.

\medskip

Note that for all $p'\geq p\geq 1$
we clearly have ($p'$SQC)
$\;\Rightarrow\;$ ($p$SQC)
$\;\Rightarrow\;$ (WQC), and that
if the quantizer $q$ of (WQC) has
finite range then ($p$SQC)
automatically holds for all $p\geq 1$.
In particular, (1SQC) amounts simply
to the requirement that there exists an appropriate scalar quantizer
$q$ with $H(q(X_1))<\infty$.

\medskip

The encoder describes $X_1^n$ with distortion
$D$ or less in two steps. First, a description
of $N_n'$ is given using Elias' code for the
integers \cite{elias}.
This takes
\be
\log N'_n + 2\log\log N'_n + O(1)
\;\;\;\;\mbox{bits}.
\label{eq:elias}
\ee
If $N'_n\leq\lfloor 2^{nb}\rfloor$,
then $N_n=N'_n$ and the above description is sufficient
for the decoder to recover a $D$-close
version of $X_1^n$ from the codebook,
so the second step is omitted.
And if $N'_n = \lfloor 2^{nb}\rfloor+1$,
then the encoder also gives
a representation of $X_1^n$ with
distortion $D$ or less using
the quantizer $q$ provided by (WQC).
This can be given in
$$\sum_{i=1}^n\lceil-\log \mu(q(X_i))\rceil
\;\;\;\;\mbox{bits}.
$$

Let $\ell_n(X_1^n)$
denote the overall description
length of the algorithm just
described. As we will see
the constant $b$ can be chosen in
such a way that
$N_n$ will eventually be small
enough so that the encoder will never
need to resort to the alternative
coding method. Therefore, in view of
(\ref{eq:elias}), to
understand this code's
compression performance
(i.e., to understand the
asymptotic behavior of
$\ell_n(X_1^n)$) it suffices
to understand the behavior of
$(\log N_n)$ for large $n$.

Suppose that a source string $X_1^n$ is
given; the probability
that any particular codeword
$Y_1^n(i)$ matches $X_1^n$ with
distortion $D$ or less is
$Q^n(B(X_1^n,D))$. If this
probability is nonzero, then,
conditional on $X_1^n$, the
distribution of $N_n$ is geometric
with parameter $Q^n(B(X_1^n,D))$. From
this observation it is easy
to deduce that $N_n$ is close to its mean,
namely $1/Q^n(B(X_1^n,D))$, when $n$
is large. The following
result is an easy consequence
of this fact and of Theorem~B below.

\medskip

{\em Theorem~A. Naive Coding Performance.}
\cite{dembo-kontoyiannis:wyner}:
Suppose that $\Xp$ is a stationary ergodic source
with first-order marginal $P_1=P$,
and that $Q$ is an arbitrary codebook
distribution on $\Ahat$ with $\Dav<\infty$.
If $D\in(\Dmin,\Dav)$ and
$\Xp$ satisfies (WQC) at $D$,
then for almost every sequence
of memoryless random codebooks
$\Cp_n$ with distribution $Q$:
\ben
\lim_{n\to\infty} \frac{1}{n}\ell_n(X_1^n) \;=\;
\lim_{n\to\infty}\frac{1}{n}\log N_n \;=\;
R(P,Q,D)
    \;\;\;\;\mbox{bits/symbol, w.p.1.}
\een
The rate-function $R(P,Q,D)$
is defined as
\be
R(P,Q,D) = \inf_W H(W\|P\times Q)\,,
\label{eq:Rdefn}
\ee
where $H(W\|V)$ denotes the relative
entropy between two probability measures
$W$ and $V$,
\ben
H(W\|V) \bydef
\left
\{ \begin{array}{ll}
   E_W[\log\frac{dW}{dV}],
           & \mbox{if the density $\frac{dW}{dV}$ exists}, \\
   \infty,      & \mbox{otherwise,}
\end{array}
\right.
\een
and the infimum in (\ref{eq:Rdefn})
is taken over all joint distributions
$W$ on $A\times\Ahat$ such that
the first marginal of $W$ is $P$
and $E_W[\rho(X,Y)]\leq D.$
This result holds as long as the
constant $b$ is chosen $b>R(P,Q,D)$.

\medskip

See Section~\ref{s:compare} for specific
examples where the asymptotic rate
$R(P,Q,D)$ can be explicitly evaluated.

\medskip

{\em Theorem~B.}
\cite{dembo-kontoyiannis:wyner}:
Let $\Xp$ be a stationary ergodic source
with first-order marginal distribution $P$,
and let $Q$ be an arbitrary
codebook distribution on $\Ahat$
with
$\Dav<\infty$. Then for all
$D\in(\Dmin,\Dav)$:
\ben
\lim_{n\to\infty}
-\frac{1}{n}\log Q^n(B(X_1^n,D)) = R(P,Q,D)
        \;\;\;\;\mbox{w.p.1}.
\een

\subsection{Entropy-Coding the Index}

Next we consider the case of entropy-coding,
where, after the codebook $\Cp_n$
has been fixed, the encoder uses
the conditional distribution
(given $\Cp_n$) of the position
of
the first $D$-close match
to optimally describe this
position to the decoder:
The truncated
index $N'_n$ is first described
using $H(N'_n|\Cp_n) + O(1)$ bits,
on the average. As before, if
$N_n\leq \lfloor 2^{nb}\rfloor$
this offers a complete $D$-close
representation of $X_1^n$. Otherwise,
the encoder adds to this an alternative
representation of $X_1^n$ using
the quantizer $q$ provided by (WQC).
On the average, this takes
$$\sum_{i=1}^nE_P\Big(\lceil-\log \mu(q(X_i))\rceil\Big)
\;\;\;\;\mbox{bits.}
$$

Let $\calL_n(X_1^n)$ denote
the overall description
length of the above coding scheme.
Next we give an upper bound on
the asymptotic
rate it achieves.
Given an arbitrary codebook
distribution $Q$ on $\Ahat,$
define the {\em output-constrained rate-distortion function}
(or lower mutual information (LMI))
\cite{zamir:index,zhang-wei:1}
by
\be
I_m(P\|Q,D)\bydef
\inf_{X\sim P,\;Y\sim Q\;E\rho(X,Y)\leq D} I(X;Y),
\label{eq:lowerMI}
\ee
where $I(X;Y)$ denotes
the mutual information between $X$ and $Y$,
and the infimum is taken over all jointly distributed
random variables $(X,Y)$ such that
$X\sim P$, $Y\sim Q,$ and $E\rho(X,Y)\leq D.$
Using the chain rule for relative entropy
it is easy to verify that the earlier
rate-function $R(P,Q,D)$ can be expressed as
\be
R(P,Q,D) =\inf_{\Qtilde}[I_m(P\|\Qtilde,D)+H(\Qtilde\|Q)],
\label{eq:achieve}
\ee
where the infimum is over all probability measures $\Qtilde$ on $\Ahat$.
As we show in Section~\ref{prelims}, the minimizer of (\ref{eq:achieve})
exists and is unique, and we denote it by $Q^*_{P,Q,D}$:
\ben
Q^*_{P,Q,D} = \argmin_{\Qtilde}[I_m(P\|\Qtilde,D)+H(\Qtilde\|Q)] .
\een

\medskip


{\em Theorem~1. Entropy-Coding Performance}:
Suppose that $\Xp$ is a stationary ergodic source
with first-order marginal distribution $P$,
and  that $Q$ is an arbitrary (i.i.d.) codebook distribution
with $\Dav<\infty$.
Assume that $D\in(\Dmin,\Dav)$ and that
$\Xp$ satisfies ($p$SQC) at $D$
for some $p>1$.
Then the rate of the entropy-coded
scheme with memoryless codebooks $\Cp_n$
with distribution $Q$, satisfies,
\be
\limsup_{n\to\infty} \frac{1}{n}E[\calL_n(X_1^n)]
\;=\;
\limsup_{n\to\infty}\frac{1}{n} H(N'_n |\Cp_n)
\;\leq\;
I_m(P\|Q^*_{P,Q,D},D)
\;\;\;\;\mbox{bits/symbol,}
\label{eq:conditional}
\ee
where
the expectation is taken over both
the message $X_1^n$ and the random codebook
$\Cp_n$. This result holds as long as $b>R(P,Q,D)$.

\medskip

We immediately obtain from this and Theorem~A above:

\medskip

{\em Corollary~1. Entropy-Coding Gain}:
Under the assumptions of Theorem~1,
the rate gain of entropy-coding over the naive
coding scheme is at least
\be
R(P,Q,D) - I_m(P\|Q^*_{P,Q,D},D) = H(Q^*_{P,Q,D} \| Q)
\;\;\;\;\mbox{bits/symbol}.
\label{eq:cor}
\ee

\medskip

As shown in \cite{zamir:index}, entropy coding
{\em without} conditioning on the codebook yields
a rate equal to the naive coding rate.
Thus, the quantity in the right hand side of
(\ref{eq:cor}) can also be thought of as the
rate-gain due to matching the entropy coder
to the specific realization of the codebook.

In Section~\ref{section:last} we generalize and refine
the bounds (\ref{eq:conditional}) and (\ref{eq:cor}).
The discussion there suggests that, as in the case
of discrete memoryless sources \cite{zamir:index},
these bounds are
tight also for general memoryless sources;
for sources with memory
the inequality (\ref{eq:conditional}) is strict
and hence the entropy-coding gain (\ref{eq:cor})
is larger.

The measure $Q^*_{P,Q,D}$ has an interesting coding interpretation
that will be clarified further in Theorem~2:
When $n$ is large, the empirical distribution of the
first matching codeword $Y_1^n(N_n)$ in the codebook
is close to $Q^*_{P,Q,D}$ with high probability.
In the case of discrete memoryless sources
this phenomenon can be explained using the method of types
as in \cite{zamir-rose:01}.
The lower mutual information $I_m(P\|\Qtilde,D)$
represents the rate achieved by a fixed-composition
codebook, namely a codebook consisting exclusively
of codewords with type $\Qtilde$. Equivalently,
$I_m(P\|\Qtilde,D)$ is the exponent in the probability
that a source string will match a type-$Q$ string
with  distortion $D$ or less. In this light,
a memoryless random codebook
with distribution $Q$ can be thought of as a union
of polynomially many fixed-composition codebooks,
where the proportion of words of type
$\Qtilde$ is $\exp[ - n H(\Qtilde\|Q) ]$.
Now, codewords of type $\Qtilde = Q$ are very frequent
in the codebook but their lower mutual information
is very high (i.e., low matching probability),
whereas codewords of type
$\Qtilde = Q^*_{P,D}$, where $Q^*_{P,D}$ achieves
the rate-distortion function (\ref{eq:R-D}),
have the lowest lower mutual information
(i.e., high matching probability),
but they are too rare in the codebook.
Therefore, we can think of the achieving
measure $\Qtilde = Q^*_{P,Q,D}$ in (\ref{eq:achieve})
as corresponding
to the codeword type that strikes the optimum
balance between the competing requirements
of high matching probability and
high frequency in the codebook.

Note that it follows from (\ref{eq:lowerMI})
and (\ref{eq:achieve}) that
\[
R(P,Q,D)  \geq I_m(P\| Q^*_{P,Q,D},D)  \geq  R(D)
\]
with equality in both inequalities if and only if
$Q$ achieves the rate-distortion function in
(\ref{eq:R-D}).

\medskip

{\em Proof outline of Theorem~1.} The equality
in (\ref{eq:conditional})
follows simply from
the observation that $E[\calL_n(X_1^n)] \geq H(N'_n |\Cp_n)$
combined with
\be
E[\calL_n(X_1^n)]
&\leq&
 H(N'_n |\Cp_n) + E\left[\sum_{i=1}^n\left\{
    [-\log \mu(q(X_i))+1]\IND_{\{N'_n=
    \lfloor 2^{nb}\rfloor+1\}}\right\}\right]\;+\;O(1)
    \nonumber\\
&\leqa&
 H(N'_n |\Cp_n) + n(M_p+1)\PR\{N_n\geq
    \lfloor 2^{nb}\rfloor+1\}\;+\;O(1)
    \nonumber\\
&\leqb&
 H(N'_n |\Cp_n) + o(n),
 \label{eq:equivalence}
\ee
where
$\IND_E$ denotes the indicator function of
an arbitrary event $E$, (a) follows by H\"{o}lder's
inequality, and (b) follows from Theorem~B.

The existence and uniqueness
of $Q^*_{PQD}$ is established
in Section~\ref{gibbs}.
The inequality in (\ref{eq:conditional})
is the main technical content of the theorem;
its proof is given in Section~\ref{proof}.
\qed

\section{The Entropy-Coding Gain}
\label{s:compare}

In this section we illustrate the gain of entropy-coding
over the naive coding scheme in two particular instances
where it can be explicitly evaluated. As mentioned in the
introduction, we first consider the case of Gaussian
codebooks with large variance. Since such a codebook
distribution is approximately uniform over the whole
real line, it is tempting to think of the entropy-coded
scheme as a ``randomized'' version of an entropy-coded,
uniform lattice quantizer. Indeed, we show that, as the
codebook variance grows to infinity, the rate achieved
by the entropy-coded scheme is at least as good as
the asymptotic rate of entropy-coded dithered
quantization (ECDQ)

Then in Section~\ref{s:flat} we consider a more general
class of approximately uniform, or ``asymptotically flat,''
codebook distributions, corresponding to appropriately
defined exponential families. In this case we
argue that the resulting compression performance
can be determined in a way analogous to the analysis
given for the Gaussian case.

\subsection{Universal Gaussian Codebooks}
\label{s:gaussian}

Let $\Xp$ be a stationary and ergodic, real-valued source
to be compressed, and suppose $\Xp$ has
zero mean $E(X_1)=0$ and finite variance
$\sigma^2=\VAR(X_1)<\infty$.
We consider memoryless random codebooks
generated according to the Gaussian
distribution $Q\sim N(0,\tau^2)$,
and we take $\rho$ be the
squared-error distortion
measure $\rho(x,y)=(x-y)^2$.
Under these assumptions,
the rate achieved by the
naive coding scheme is \cite{dembo-kontoyiannis:wyner},
\be
R(P,Q,D) = \left\{ \begin{array}{ll}
    \infty\,,   & \;\;\;\;D=0\\
    \frac{1}{2}\log\left(\frac{v}{D}\right)
      -(\log e)\frac{(v-D)(v-\sigma^2)}
        {2v\tau^2}\,,
            & \;\;\;\;0<D<\sigma^2+\tau^2\\
    0\,,        & \;\;\;\;D\geq \sigma^2+\tau^2,
 \end{array} \right.
\label{RpqdGauss}
\ee
where
$$v\bydef\frac{1}{2}\left[\tau^2+\sqrt{\tau^4+4D\sigma^2}\right].$$
[Note that here $\Dmin=0$, $\Dav=\sigma^2+\tau^2$.]
We observe that in this case $R(P,Q,D)$ depends only
on the first and second moments of the source
distribution, and that asymptotically for large
codebook variance it takes the form
\be
R(P,Q,D) = \frac{1}{2}\log\left(\frac{\tau^2}{e D}\right) + o(1)
\label{eq:naive-as}
\ee
where $o(1) \rightarrow 0$ as $\tau^2 \rightarrow \infty$.

\medskip

{\em Remark:}
In more familiar information-theoretic
terms, the rate-function $R(P,Q,D)$
can equivalently be expressed as
\be
R(P,Q,D) = \inf_{(X,Y)}\,[I(X;Y)
    +H(Q_Y\|Q)]
\label{eq:alternative}
\ee
where the infimum is over
all jointly distributed random variables
$(X,Y)$ with values in $A\times\Ahat$,
such that $X$ has distribution $P$,
$E[\rho(X,Y)]\leq D$, and $Q_Y$ denotes
the distribution of $Y$;
cf. \cite{yang-kieffer:1}.

This expression shows that,
typically, the rate achieved by the naive coding
scheme is strictly suboptimal, unless of course
the source itself is memoryless and $Q$ is chosen
to minimize $R(P,Q,D)$. In fact
from (\ref{eq:alternative}) it is immediate that
for a memoryless source with rate-distortion
function $R(D)$ we indeed have
\be
R(D)=\inf_{\Qtilde} R(P,\Qtilde,D),
\label{eq:R-D}
\ee
where the infimum is over all
probability distributions $\Qtilde$ on $\Ahat$.

\medskip

{\em Example. Known Variance. }
Now suppose that the source $\Xp$ is believed
to be $\iid$ Gaussian
with $N(0,\sigma^2)$ distribution.
As is well-known
\cite{berger:book,cover:book},
for any $D\in(0,\sigma^2)$
the optimal coding distribution
is $Q^*\sim N(0,\sigma^2-D)$,
therefore we construct memoryless
random codebooks according to $Q^*$.
But instead of the Gaussian source we expected,
we are faced with data from some arbitrary
stationary ergodic $\Xp$ with zero mean
and variance $\sigma^2$. From
the previous example it follows
that the asymptotic rate achieved by the
naive coding scheme will be
[substituting $\tau^2 = \sigma^2-D$ in
(\ref{RpqdGauss})]
$$\frac{1}{2}\log\left(\frac{\sigma^2}{D}\right)
\;\;\;\;\mbox{bits/symbol.}$$
This is exactly the rate-distortion
function of the $\iid$ $N(0,\sigma^2)$
source, so
the rate achieved is
the same as what we would have
obtained on the Gaussian source we
originally expected.
This coincides with Sakrison's robust
fixed-rate for a class of sources
\cite{sakrison:69}.
It is yet another version of the
folk theorem that the Gaussian
source is the hardest one to compress,
among all real-valued sources with
a fixed variance;
cf. \cite{lapidoth:97}.

\medskip

Turning back to the general case, suppose
$\Xp$ is a zero-mean,
stationary and ergodic, real-valued source,
with variance $\sigma^2=\VAR(X_1)<\infty$,
and let the codebook distribution be
$Q\sim N(0,\tau^2)$.
Choose and fix a distortion level $D>0$. From
(\ref{eq:naive-as}) we have that,
for $\tau^2$ large,
the rate achieved by the naive coding scheme is
$$\log (\tau) + O(1)\;\;\;\;\mbox{bits/symbol}$$
which of course grows to infinity as
$\tau^2\to\infty$. On the other hand,
as we show in the following proposition
the rate achieved by the entropy-coding scheme
stays bounded, and for memoryless sources
it coincides with the asymptotic (large vector
dimension) rate of entropy-coded dithered lattice
quantization (ECDQ). This confirms the natural
intuition that the behavior of a random codebook
with an approximately flat distribution should
mimic the behavior of an entropy-coded
uniform quantizer.

\medskip

{\em Proposition~1.
Entropy-Coding Gain for Universal Gaussian Codebooks}:
Let $\Xp$ be a real-valued, zero-mean,
stationary ergodic source, with finite variance
$\sigma^2=\VAR(X_1),$ let $D>0$ be a fixed
distortion level, and take $Q\sim N(0,\tau^2)$.
Let $Q^*_{PQD}$ be as in Theorem~1. We have:
\begin{itemize}
\item[(i)]
The measure $Q^*_{PQD}$ converges to $P*N(0,D)$
as $\tau\to\infty$, in that $Q^*_{PQD}$
    has a density $f_{Q^*_{PQD}}(y)$
    (with respect to Lebesgue measure)
    for all $\tau$ large enough and
    $$f_{Q^*_{PQD}}(y)\to E_P[\phi_D(y-X)]\;\;\;\;
    \mbox{as $\tau^2\to\infty,$ for all $y\in\RL$,}$$
    where $\phi_D$ denotes the density of
    the $N(0,D)$ distribution.
\item[(ii)]
The upper bound $I_m(P\|Q^*_{PQD},D)$ to the
    rate achieved by the entropy-coding scheme
    satisfies
    \ben
    \lim_{\tau^2\to\infty} I_m(P\|Q^*_{PQD},D) = I(X;X+Z_D)
    \;\;\;\;\mbox{bits/symbol,}
    \een
    where $X\sim P$, and $Z_D$ denotes a $N(0,D)$ random
    variable independent of $X$.
\item[(iii)]
    If the source $\Xp$ is memoryless then
    as $\tau\to\infty$ the rate achieved by the
    entropy-coding scheme is no greater than
    \ben
    R(D)+\frac{1}{2}
    \;\;\;\;\mbox{bits/symbol,}
    \een
    where $R(D)$ is the rate-distortion function of $\Xp$.
\end{itemize}

\medskip

A proof outline for Proposition~1 is given in Appendix~A.
As we will discuss in Section~\ref{section:last},
for sources with memory the entropy-coding gain is generally
significantly larger. In fact when $\Xp$ is not memoryless:
\begin{enumerate}
\item
the rate achieved by the entropy-coding scheme is
actually equal to the mutual information rate
$I(\Xp; \Xp +\Zp_D)$, where $\Zp_D$ is a white
Gaussian process with variance $D$;
\item the result in part~(iii) of the proposition
is valid for all stationary and ergodic sources.
\end{enumerate}

\subsection{Approximately Flat Codebooks and Difference Distortion Measures}
\label{s:flat}
We now extend the asymptotic result above to more general
codebook distributions and to arbitrary difference distortion
measures (not necessarily squared error).

Suppose that $\hat{A} = A=\RL$ and that
$\rho$
is a difference distortion measure of the
form $\rho(x,y) = \rho(y-x)$ for some
$\rho:\RL\to[0,\infty).$
Here we consider real-valued,
stationary ergodic sources $\Xp$, and codebook
distributions $Q$ that have a density $f_Q$
(with respect to Lebesgue measure).

We begin by deriving a lower bound for the
rate-function $R(P,Q,D)$, in the spirit of the Shannon Lower Bound
\cite{linder-zamir:94}.

\medskip

{\em Lemma~SLB. A ``Shannon Lower Bound'' for Difference Distortion Measures}:
Assume that the codebook distribution $Q$
has a density $f_Q$ (with respect to Lebesgue measure),
and let $\Qmax = \sup_y f_Q(y)$.
Then,
\begin{equation}
\label{SLB}
   R(P,Q,D) \geq  \log(1/\Qmax) - \hmax(D),
\end{equation}
where  $\hmax(D)$ is the maximum entropy associated with $\rho$ and $D$,
\ben
\label{max_ent}
\hmax(D) = \max_{f: \ E_f [ \rho(Z) ] \leq D } h(f)
\een
and where $h(f) = - \int f(x) \log f(x) dx = h(Z)$
denotes the differential entropy of a random variable
$Z$ with density $f$.

\medskip

Note that the lower bound (\ref{SLB})
is independent of the source distribution $P$.

\medskip

{\em Proof.}
Consider the infimum in (\ref{eq:alternative}).
For any jointly distributed $(X,Y)$ such that
$E[\rho(Y-X)]\leq D$, let $Q_Y$ and $f_Y$ denote the
measure and the density describing the distribution
of $Y$, respectively. We can then write \cite{cover:book},
\[
I(X;Y) = h(Y) - h(Y|X) = h(Y) - h(Y-X|X) \geq h(Y) - h(Y-X)
\geq h(f_Y) - \hmax(D)
\]
where the first inequality holds since conditioning reduces
the entropy.
On the other hand, we can expand
$$H(Q_Y\|Q) = - h(f_Y) + E_{Q_Y} [ - \log f_Q(Y) ]
\geq
-h(f_Y) + \log(1/\Qmax).$$
Combining, we get
\be
  I(X;Y) + H(Q_Y\|Q) \geq \log(1/\Qmax) - \hmax(D).
\label{eq:lb}
\ee
Now note that we have implicitly assumed that $Y$ has
a conditional density given $x$ for $P$-almost all $x$,
but if it did not then the relative entropy
$H(W\|P\times Q)=E_P[H(W(\cdot|X)\|Q)]=I(X;Y) + H(Q_Y\|Q)$
between
the joint distribution $W$ of $(X,Y)$ and $(P\times Q)$
would be infinite, so the above bound would still
hold. Therefore, in view
of (\ref{eq:alternative}),
the right-hand side of (\ref{eq:lb}) is
also a bound for $R(P,Q,D)$.
\qed

\medskip

Similarly to the Shannon lower bound for the rate-distortion
function \cite{linder-zamir:94},
the lower bound above turns out to be tight for
several interesting special cases.
To see this we first derive an upper bound for $R(P,Q,D)$.
In (\ref{eq:alternative})
we can always pick
\begin{equation}
\label{additive-noise-channel}
Y = X + Z_D,
\end{equation}
where $Z_D \sim f_D$ is independent of
$X$ and it achieves the maximum entropy associated with $\rho$ and $D$,
i.e., $h(Z_D) = h(f_D) = \hmax(D)$.
For this choice
$I(X;Y) = h(Y) - h(Z_D) = h(f_Y) - \hmax(D)$,
where $f_Y$ is the density of $Y=X+Z_D$.
Therefore,
\begin{eqnarray}
R(P,Q,D) & \leq &
h(f_Y) - \hmax(D) + H(Q_Y \| Q)   \\
\label{upper}
& = & E_{f_Y} [ - \log f_Q(Y) ]  -  \hmax(D),
\end{eqnarray}
where $Y = X + Z_D$.

Now if $f_Q(y)$ is continuous near $\ymax = \argmax_y f_Q(y)$,
and $f_Y$ is concentrated around $\ymax$,
then  $E_{f_Y} [ - \log f_Q(Y) ] \approx \log(1/\Qmax)$,
and the two bounds (\ref{SLB}) and (\ref{upper}) are close.
Since the lower bound (\ref{SLB}) is independent of $P$,
closeness of the bounds would imply that $R(P,Q,D)$
is only weakly dependent on the source distribution.
For example, for a uniform codebook distribution
$Q \sim U[-K,K]$ we have $\Qmax = 1/2K$
so the lower bound is
$R(P,Q,D) \geq \log(2K) - \hmax(D)$.
On the other hand,
if $K$ is large enough so that
$f_Y(y)=0$ for $|y|>K$, then
$E_{f_Y} [ - \log f_Q(Y) ] = \log(2K)$,
and the lower bound is tight.
See also \cite[Lemma~1]{zamir:index}

More generally, suppose that the codebook
distribution $Q=Q_s$ has an exponential density of the form
\begin{equation}
\label{maxentQ}
     f_s(y) = B_s \; \exp( - s g(y) )  \ , \  s>0,
\end{equation}
where $g$ is any suitable (nonnegative) function with $g(0)=0$.
Gaussian codebooks correspond to the case $g(y) = y^2$,
while uniform codebooks correspond to a ``well-shaped'' $g$.
Moreover, for any ``nice'' $g$ (as stated rigorously in
the next lemma), as $s \rightarrow 0$ the exponential density
$f_s(y)$ tends to be {\em locally uniform}
relative to the $Y$ of (\ref{additive-noise-channel}).
This explains the following asymptotic characterization
of $R(P,Q,D)$.

\medskip

{\em Lemma~TIGHT. Asymptotically Flat Codebooks}:
For a difference distortion measure and an exponential codebook
distribution $Q = Q_s$ of the form (\ref{maxentQ}),
if $E[ g(X+Z_D) ]$ is finite, then the lower bound (\ref{SLB})
becomes tight as $s\to 0$:
\[
 R(P,Q_s,D) = \log(1/B_s) - \hmax(D) + o(1).
\]

{\em Proof.}
For $Q=Q_s$ we have $\Qmax = B_s$ and $\ymax = 0$,
so the lower bound (\ref{SLB}) is equal to
$\log(1/B_s) - \hmax(D)$.
On the other hand, $- \log f_Q(y) = \log(1/B_s) + s g(y)$,
so the upper bound (\ref{upper}) is equal to
$\log(1/B_s) + s E[ g(X+Z_D) ] -  \hmax(D)$,
which approaches the lower bound as $s \rightarrow 0$.
\qed

\medskip

An interesting consequence of Lemma~TIGHT
is that, like for uniform codebooks, for very
flat codebook distributions $Q_s$
the rate-function $R(P,Q_s,D)$
is {\em almost independent} of the source distribution $P$.
In particular, for a  Gaussian $Q\sim N(0,\tau^2)$
and any source $P$,
\[
R(P,Q,D) = \frac{1}{2}\log(2 \pi \tau^2) - \hmax(D) + o(1)
\]
as $\tau \to \infty$.
If $\rho$ = squared error, then
$\hmax(D)= (1/2)\log(2 \pi e D)$, and we obtain
$R(P,Q,D) = (1/2)\log(\tau^2/ e D) + o(1)$
as in (\ref{RpqdGauss}).

Another consequence of the asymptotic tightness of the upper bound
(\ref{upper}) is that an additive maximum entropy noise channel
of the form (\ref{additive-noise-channel})
asymptotically achieves the minimizations
(\ref{eq:Rdefn}) and (\ref{eq:alternative}).
This observation extends the asymptotic additive Gaussian noise
channel characterization of $I_m(P\|Q^*_{P Q D},D)$ in Proposition~1.
We state this result in the following proposition and
prove it in Appendix~D.

\medskip

{\em Proposition~2. Entropy-Coding Gain for Approximately Flat Codebooks}:
Let $\rho$ be a difference distortion measure
such that the maximum entropy $\hmax(D)$ defined in (\ref{max_ent})
exists and is strictly monotonically increasing with $D$.
Let $Q_s$ be any exponential codebook distribution of the form (\ref{maxentQ})
such that $E[ g(X+Z_D) ]$ is finite.
Then:
\begin{itemize}
\item[(i)]
The measures $Q^*_{P Q_s D}$ converge to $P * f_D$ as $s\to 0$,
where $P * f_D$ is the distribution of $Y = X + Z_D$,
and where $Z_D$ is the random variable achieving $\hmax(D)$.
This convergence is in the sense that the density
of $Q^*_{P Q_s D}$ converges
to the density of $Y = X + Z_D$.
\item[(ii)]
The upper bound $I_m(P\|Q^*_{PQ_sD},D)$ to the
    rate achieved by the entropy-coding scheme
    satisfies
    \ben
    \lim_{s\to 0} I_m(P\|Q^*_{PQ_sD},D) = I(X;X+Z_D)
    \;\;\;\;\mbox{bits/symbol,}
    \een
    where $X\sim P$, and $Z_D$ is the maximum entropy
    random variable achieving $\hmax(D)$.
\item[(iii)]
    If the source $\Xp$ is memoryless, then as
$s \rightarrow 0$ the rate achieved by the entropy-coding
scheme is no greater than
\begin{equation}
\label{Cbound}
R(D) + C^* \;\;\;\;\mbox{bits/symbol},
\end{equation}
where the universal constant $C^*$ is defined as
\[
C^* = C^*(\rho, D) = \sup_{U: \; E\rho(U) \leq D} I(U; U+ Z_D) .
\]
\end{itemize}

Note that $C^*$ is an upper bound for the ``min-max capacity''
defined in \cite{zamirWZ} in connection with the Wyner-Ziv
problem. In particular, for an $r$th-power distortion measure
$\rho(\hat{x}-x) = |\hat{x}-x|^r$, we have
$0.5 \leq C^* \leq 1$ bit,  and for $r=2$ we have $C^* = 1/2$
bit in accordance with Proposition~1~(iii);
see \cite{feder-zamir:92}.

We note that the codebook distribution $Q_s$
in Proposition~2 is independent of the source,
of the distortion measure, {\em and} of the distortion level.
This implies a strong robustness property for
a memoryless codebook drawn from $Q_s$:
For sufficiently small $s$, such a codebook
is universal (in the sense of the bounded loss in (\ref{Cbound}))
for any source and any distortion criterion
admissible by the proposition.
Thus, we may imagine that we first fix the codebook; then we select
the desired distortion criterion to generate the $D$-balls;
and finally we let the source induce the codeword distribution which
determines the entropy code.


\section{An Almost Sure Conditional Limit Theorem}
\label{gibbs}

\subsection{Preliminaries}
\label{prelims}
As before, we assume throughout this section
that $\Xp$ is a stationary ergodic source
and $Q$ is an arbitrary codebook distribution
with $0\leq \Dmin<\Dav<\infty$. Also we fix
a distortion level $D\in(\Dmin,\Dav)$. Under
these conditions, from
\cite[Theorem~2]{dembo-kontoyiannis:wyner}
we know that $R(P,Q,D)$ is finite and strictly positive,
and that the infimum in its definition
in (\ref{eq:Rdefn}) is always achieved by some joint
distribution $W^*$ with $E_{W^*}[\rho(X,Y)]=D.$
Moreover, since the set of $W$ over which
the infimum is taken is convex, from
\cite{csiszar:75} we know that this $W^*$
is the unique minimizer.

Alternatively, $R(P,Q,D)$ can be expressed as
\be
(\ln 2)R(P,Q,D)=\sup_{\la\leq 0}[\la D -\LA(\la)]
    =\la^*D-\LA(\la^*),
\label{eq:represent}
\ee
where
$$\LA(\la)\bydef E_P\left[\ln E_Q \left(e^{\la\rho(X,Y)}\right)\right],$$
and $\la^*$ is the unique negative real
number with
\be
\LA'(\la^*)=D;
\label{eq:condition}
\ee
where prime denotes derivative,
cf. \cite[Theorem~2]{dembo-kontoyiannis:wyner}.

Let $Q'=W_Y^*$ denote the $Y$-marginal of $W^*$. From
the above discussion and from equations
(\ref{eq:alternative}), (\ref{eq:lowerMI})
and (\ref{eq:achieve}) it follows
that the infimum in (\ref{eq:achieve}) is
uniquely achieved by $Q'$.
That is
$$ Q' = W_Y^* = Q^*_{P,Q,D} $$
where as before $P = P_1$ denotes the first-order
marginal of the source
distribution.
Now let
$\Qhatn$ denote the empirical distribution
induced by the matching codeword
$Z_1^n\bydef Y_1^n(N_n)$ on $\Ahat$:
\[
\Qhatn \bydef \hat{P}_{Y_1^n(N_n)}  \bydef \frac{1}{n}\sum_{i=1}^n\delta_{Z_i}.
\]

\subsection{Results}
Our first result says that,
when $n$ is large,
$\Qhatn\approx Q'$
with high probability.
This generalizes and strengthens
the ``favorite-type theorem''
of \cite{zamir-rose:01}.

\medskip

{\em Theorem~2. Empirical Distribution of the Matching Codeword (``Favorite Type'')}:
\begin{itemize}
    \item[(i)]
For every (measurable)
$E\subset\Ahat$,
any $\delta>0$,
and $\BBP$-almost
every source realization $x_1^\infty$, as $n\to\infty$
we have:
$$\Pr \left\{|\Qhatn(E) - Q'(E)|>\delta\,\Big| \, X_1^n=x_1^n \right\}\to 0
\;\;\mbox{exponentially fast.}$$
    \item[(ii)]
With probability one,
$$\Qhat_n\Rightarrow Q',\;\;\;\;\mbox{as $n\to\infty$,}$$
where `$\Rightarrow$' denotes weak convergence
of probability measures.
\end{itemize}

\medskip

As we will see below, Theorem~2 is a
consequence of the following generalization
of the conditional limit theorem
(see, e.g., \cite[Ch.12]{cover:book} for the
standard form of the conditional
limit theorem).

\medskip

{\em Theorem~3. Almost Sure Conditional Limit Theorem}:
Let $X_1^n$ and $Y_1^n$ be two independent random
vectors with distributions $P_n$ and $Q^n$, respectively,
and write $\Phatn = \hat{P}_{Y_1^n}$ for the empirical
distribution of $Y_1^n$. For every
(measurable) $E\subset\Ahat$,
any $\delta>0$,
and for $\BBP$-almost every realization $x_1^\infty$,
as $n\to\infty$ we have:
\ben
\Pr\left\{|\Phatn(E) - Q'(E)|>\delta\,\Big |\,
\rho_n(X_1^n,Y_1^n)\leq D\;\;\mbox{and}\;\;
X_1^n=x_1^n\right\}\to 0
\;\;\mbox{exponentially fast.}
\label{eq:thm3}
\een

\medskip

Note that since $\rho_n(X_1^n,Y_1^n)\leq D$ is a rare event,
the conditional probability in Theorem~3 would have been different
without conditioning on a $\BBP$-almost sure realization $x_1^\infty$.
We first deduce Theorem~2 from Theorem~3 and then
we give the proof of Theorem~3.

\medskip

{\em Proof of Theorem~2.}
Observe that
\ben
\lefteqn{
        \Pr\left\{|\Qhatn(E) - Q'(E)|>\delta
        \, \Big| \,X_1^n\right\}
    }\\
& &\eqa
    \sum_{k\geq 1}
    \Pr \left\{|\Qhatn(E) - Q'(E)|>\delta\, \Big|
    N_n=k\;\;\mbox{and}\;\;X_1^n \right\}\,  \PR\{N_n=k\,|\,X_1^n\}\\
& &\eqb
    \sum_{k\geq 1}
    \Pr\left\{|\Phatn(E) - Q'(E)|>\delta\,\Big |\,
    \rho_n(X_1^n,Y_1^n)\leq D\;\;\mbox{and}\;\;
    X_1^n\right\}\PR\{N_n=k\,|\,X_1^n\}\\
& &\eqc
    \Pr\left\{|\Phatn(E) - Q'(E)|>\delta\,\Big |\,
        \rho_n(X_1^n,Y_1^n)\leq D\;\;\mbox{and}\;\;
        X_1^n\right\},
\een
where $(a)$ and $(c)$ follow from the fact that
$N_n<\infty$, eventually with probability one
(by Theorem~A);
and $(b)$ follows from the observation that,
due to the codewords' independence,
the random variables $\rho_n(X_1^n,Y_1^n(k)), \; k=1,2,\ldots$
are conditionally independent given $X_1^n$.
This implies that, given $N_n$,
the distribution of the matching codeword is
exactly the same as the distribution of $Y_1^n$
conditioned on the event $\{\rho_n(X_1^n,Y_1^n)\leq D\}$.

This together with Theorem~3 proves~(i).  From~(i)
and the Borel-Cantelli lemma we
conclude that, for any measurable
$E\subset\Ahat$,
$$\Qhatn(E)\to Q'(E),\;\;\;\;\mbox{as $n\to\infty$,}
\;\;\mbox{w.p.1.}$$
Since $\Ahat$ is a Polish space,
there exists a countable
convergence-determining class
${\cal E}=\{E_i\}\subset\hat{\cal{A}}$.\footnote{For example,
let $B$ be a countable dense subset of
$\Ahat$, and take $\cal E$ to be the
collection of all open
balls with rational radii
centered at the points of $B$,
together with all finite intersections
of such balls.
Then $\cal E$ is countable (by construction)
and it is easy to check that
\cite[Theorem~2.3]{billingsley:cpm} applies,
verifying that
$\cal E$ is convergence-determining.}
Therefore,
with probability one we have that
$$\Qhatn(E_i)\to Q'(E_i),
\;\;\;\;\mbox{as $n\to\infty$,}\;\;
\mbox{for all $i$,}$$
and this implies $(ii)$.
\qed

\medskip

{\em Proof of Theorem~3.}
The probability in
Theorem~3
can be expanded as
\ben
\lefteqn{
    \Pr\left\{|\Phatn(E) - Q'(E)|>\delta
    \;\;\mbox{and}\;\;\rho_n(X_1^n,Y_1^n)\leq D
    \,\Big |\, X_1^n\right\}
    \Big/
    \Pr\{\rho_n(X_1^n,Y_1^n)\leq D\,|\, X_1^n\}
    }\\
& & =\;
    \Pr\left\{\Phatn(E) < Q'(E)-\delta
        \;\;\mbox{and}\;\;\rho_n(X_1^n,Y_1^n)\leq D
        \,\Big |\, X_1^n\right\}
        \Big/ Q^n(B(X_1^n,D))\\
& & \hspace{0.4in}
    +\;
    \Pr\left\{\Phatn(E) > Q'(E)+\delta
        \;\;\mbox{and}\;\;\rho_n(X_1^n,Y_1^n)\leq D
        \,\Big |\, X_1^n\right\}
        \Big/ Q^n(B(X_1^n,D)).
\een
We only treat the first of the two terms above;
the second one can be dealt with similarly.

If $Q'(E)\leq \delta$
there is nothing to prove, so let us assume that
$Q'(E)>\delta$.
In view of Theorem~B, it suffices to show that
\be
\limsup_{n\to\infty}\frac{1}{n}\log
    \Pr\left\{\Phatn(E) < Q'(E)-\delta
        \;\;\mbox{and}\;\;\rho_n(X_1^n,Y_1^n)\leq D
        \,\Big |\, X_1^n\right\}
    < - R(P,Q,D)
    \;\;\;\;\mbox{w.p.1.}
\label{eq:thm3target}
\ee
This will be proved by an application
of the G\"{a}rtner-Ellis theorem. Toward
that end,
choose and fix an arbitrary realization
$x_1^\infty$ of $\Xp$, and define a
sequence of random vectors $\{\xi_n\}$
in $\RL^2$ as
$$\xi_n\bydef \left(\rho_n(x_1^n,Y_1^n),\Phatn(E)\right)
=\frac{1}{n}\sum_{i=1}^n\left(\rho(x_i,Y_i),\IND_{E}(Y_i)\right),$$
where the random variables $\{Y_i\}$
are as in the statement of the theorem.
Let $\THE_n(\la)$ denote the log-moment generating
function of $\xi_n$,
$$\THE_n(\la)\bydef\ln
    E_{Q^n}\left[
        \exp\{\la_1\rho_n(x_1^n,Y_1^n)+\la_2\Phatn(E)\}
    \right],
    \;\;\;\;\;\;
    \la=(\la_1,\la_2)\in(-\infty,0]^2.$$
By the ergodic theorem we have, for $\BBP$-almost
every realization $x_1^\infty$,
\be
    \lim_{n\to\infty} \frac{1}{n}\THE_n(n\la)
&=&
    \lim_{n\to\infty}
    \frac{1}{n}\sum_{i=1}^n\ln
    E_Q\left[
        \exp\{\la_1\rho(x_i,Y)+\la_2\IND_{E}(Y)\}
        \right]
    \nonumber\\
&=&
\THE(\la)
\;\bydef\;
E_P\left\{\ln E_Q\left[
                \exp\{\la_1\rho(X,Y)+\la_2\IND_{E}(Y)\}
        \right]
    \right\},
    \label{eq:logmg}
\ee
where, by Jensen's inequality,
the limiting log-moment generating
function $\THE(\la)$ satisfies
$-\infty<(\la_1\Dav+\la_2)
\leq\THE(\la)\leq 0$. It is
easy to check (using the
dominated convergence
theorem) that $\THE(\la)$
is differentiable, with
partial derivatives
\ben
\frac{\partial\THE}{\partial\la_1}
&=&
    E_{W^{(\la)}}[\rho(X,Y)],\\
\frac{\partial\THE}{\partial\la_2}
&=&
        W^{(\la)}_Y(E),
\een
where $W^{(\la)}$ is the probability measure
defined on $A\times\Ahat$
by
\ben
\frac{dW^{(\la)}(x,y)}{d(P\times Q)}
\bydef
\frac{\exp\{\la_1\rho(x,y)+\la_2\IND_{E}(y)\}}
{E_Q\left[\exp\{\la_1\rho(x,Y)+\la_2\IND_{E}(Y)\}
    \right]},
\een
and $W^{(\la)}_Y$ is the $Y$-marginal of $W^{(\la)}$.

Note that $\THE(\la)$ is a convex
function, and define its convex dual,
$$\THE^*(z)=\sup_{\la_1\leq 0,\;
        \la_2\leq 0}[(z,\la)-\THE(\la)],
        \;\;\;\;z=(z_1,z_2)\in\RL^2,$$
where $(z,\la)$ denotes the usual Euclidean inner
product $(z,\la)=z_1\la_1+z_2\la_2$.

In view of (\ref{eq:logmg}) and of
the above discussion, we can apply
the G\"{a}rtner-Ellis theorem
\cite[Theorem~2.3.6]{dembo-zeitouni:book}
to conclude that, with probability one,
the probabilities in
(\ref{eq:thm3target}) satisfy
\ben
\limsup_{n\to\infty}\frac{1}{n}\ln
        \Pr\left\{\Phatn(E) < Q'(E)-\delta
        \;\;\mbox{and}\;\;\rho_n(X_1^n,Y_1^n)\leq D
        \,\Big |\, X_1^n\right\}
        \leq -\inf_{z_1\in[0,D],\,z_2\in[0,q]} \THE^*(z_1,z_2),
\een
with probability one,
where we write $q=Q'(E)-\delta>0$.  From
its definition, it is obvious that
$\THE^*(z_1,z_2)$ is nonincreasing in each of
its coordinates. Therefore, to prove (\ref{eq:thm3target})
and conclude the proof of the theorem it
suffices to show (recall (\ref{eq:represent})) that:
\be
\THE^*(D,q)>(\ln 2)R(P,Q,D)=\la^*D-\LA(\la^*).
\label{eq:thm3targetb}
\ee

To prove (\ref{eq:thm3targetb}) we consider
$$
g(\la_1,\la_2)\bydef\la_1 D+\la_2 q-\THE(\la_1,\la_2).
$$
Using the dominated convergence theorem as before
we can differentiate $g$ with respect to
$\la_2$ to get that
for all $(\la_1,\la_2)\in(-\infty,0]^2$,
$$
\frac{\partial g(\la_1,\la_2)}{\partial\la_2} =
q-E_{P\times Q}\left[\IND_{E}(Y)\frac{
\exp\{\la_1\rho(X,Y)+\la_2\IND_{E}(Y)\}}
{E_Q[ \exp\{\la_1\rho(X,Y')+\la_2\IND_{E}(Y')\}
]
}\right]=
q-W^{(\la)}_Y(E     ),$$
where at the endpoint $\la_2=0$ this is
understood as the corresponding right-derivative.
Now, if this derivative evaluated at
$(\la_1,\la_2)=(\la^*,0)$ is nonnegative,
i.e., if
\be
W^{(\la^*,0)}_Y(E      )\leq q <Q'(E      ),
\label{eq:compare}
\ee
then, after some simple algebra,
$W^{(\la^*,0)}$ is easily seen to
satisfy
$$(\ln 2)H(W^{(\la^*,0)}\|P\times Q)=\la^* D-\LA(\la^*)=(\ln 2)R(P,Q,D),$$
and also
$$E_{W^{(\la^*,0)}}[\rho(X,Y)]=\LA'(\la^*)=D$$
(see (\ref{eq:condition})).
Since as we remarked above $R(P,Q,D)$ is
uniquely achieved by $W^*,$ we must have that
$W^{(\la^*,0)}=W^*$, and, in particular, $W^{(\la^*,0)}_Y=W_Y^*=Q'$.
But this contradicts (\ref{eq:compare}).

Therefore, it must be the case that
$$\left.\frac{\partial g(\la_1,\la_2)}
{\partial\la_2}\right|_{(\la^*,0)}<0,$$
which means that by taking $\la_2=\la'$ slightly
negative, we can make $g(\la^*,\la')$ strictly
larger than $g(\la^*,0)=(\ln 2)R(P,Q,D)$. Hence
$$\THE^*(D,q)=\sup_{\la_1,\,\la_2} g(\la_1,\la_2)
\geq g(\la^*,\la')>(\ln 2)R(P,Q,D),$$
establishing
(\ref{eq:thm3targetb})
and thereby completing the proof.
\qed

\section{Entropy-Coding Performance}
\label{proof}


Before giving the proof of Theorem~1 we need to state
four simple Lemmas that establish some of the technical
properties we need in the proof.

\subsection{Four Lemmas}

Recall the notation and assumptions of
Section~\ref{prelims}.
We begin with some preliminary lemmas.

\medskip

{\em Lemma~1}:
$$\lim_{\delta\downarrow 0}\;
\sup_{\{F_i\}}\;
\inf_{\Qtilde: |\Qtilde-Q'|<\delta}\;
H(\Qtilde\|Q)=H(Q'\|Q),$$
where the supremum is taken over all
finite partitions $\{F_1,F_2,\ldots,F_k\}$
of $\Ahat$, and for any such partition
the infimum is over all probability
measures $\Qtilde$ on $\Ahat$
such that $|\Qtilde(F_i)-Q'(F_i)|<\delta$
for all $i$.

\medskip

{\em Proof.} Since $Q'$ is always among the measures
over which the infimum is taken, we obviously have
that the above left-hand side is no larger than the
right-hand side.

To prove the corresponding lower bound let
$\epsilon>0$ arbitrary, and choose a finite
partition ${\cal P}=\{F_1,\ldots,F_k\}$
such that $H(Q'_{\cal P}\|Q_{\cal P})>
H(Q'\|Q)-\epsilon,$ where for any measure
$\mu$ and any partition $\cal P$ we write
$\mu_{\cal P}$ for the corresponding
discrete measure which assigns
probability $\mu(F_i)$ to each
$i$ in the alphabet $\{1,2,\ldots,k\}$.
The fact that this is possible follows from
\cite[Chapter~2]{pinsker:book}
and the fact that $H(Q'\|Q)\leq R(P,Q,D)<\infty$.
Without loss of generality we assume that
$Q(F_i)\neq 0$ for all $F_i\in{\cal P}$.

By the uniform continuity of relative
entropy on a finite alphabet,
we can choose $\delta_0>0$ small enough
so that
$$H(\Qtilde_{\cal P}\|Q_{\cal P})>
H(Q'_{\cal P}\|Q_{\cal P})-\epsilon$$
for all probability measures $\Qtilde$
on $\{1,\ldots,k\}$ with
$|\Qtilde_{\cal P}(F_i)-Q_{\cal P}(F_i)|<\delta_0$
for all $i$.
Then by the data processing
inequality for relative
entropy,
for all $\delta<\delta_0$,
$$
\inf_{\Qtilde: |\Qtilde-Q'|<\delta_0}\;
H(\Qtilde_{\cal P}\|Q_{\cal P})
\geq
H(Q'_{\cal P}\|Q_{\cal P})-\epsilon
\geq
H(Q'\|Q)-2\epsilon,
$$
and since $\epsilon>0$ was
arbitrary, we are done.
\qed

\medskip

The following lemma contains
a simple observation based on
Lemma~1; it is stated here
without proof.

\medskip

{\em Lemma~2}:
For any $\epsilon>0$ there is a $\delta>0$
and a finite partition ${\cal P}=\{F_1,\ldots,F_k\}$
of $\Ahat$ such that
$$
R(\delta)\bydef
\inf_{\Qtilde: |\Qtilde-Q'|<\delta}\;
H(\Qtilde\|Q)>H(Q'\|Q)-\epsilon.$$
Writing $I(\delta)\bydef R(P,Q,D)-R(\delta)$,
we have
$$I_m(P\|Q',D)\leq I(\delta)\leq I_m(P\|Q',D)+\epsilon.$$

\medskip

Given
a $\delta>0$
and a
finite partition ${\cal P}=\{F_1,\ldots,F_k\},$
let $B_\delta$ denote the set of probability
measures on  $\Ahat$ that are $\delta$-close to
$Q'$ on the sets $F_i$:
\be
B_\delta\bydef\{\Qtilde\,:\,|\Qtilde(F_i)-Q'(F_i)|<\delta
\;\;\mbox{for all}\;i\}.
\label{eq:bdelta}
\ee

{\em Lemma~3.}
Let $X_1^n$ and $Y_1^n$ be two independent random
vectors with distributions $P_n$ and $Q^n$, respectively,
and write $\Phatn$ for the empirical
distribution of $Y_1^n$. Suppose
a $\delta>0$
and a
finite partition ${\cal P}=\{F_1,\ldots,F_k\}$
of $\Ahat$ are given, and write
$$p_n\bydef p_n(x_1^n,\delta)\bydef
\PR\{\rho_n(x_1^n,Y_1^n)\leq D\,|\,\Phatn\in B_\delta \}.$$
Then we have:
\ben
\lim_{n\to\infty}-\frac{1}{n}\log p_n(X_1^n,\delta) =
I(\delta)
\;\;\;\;\mbox{w.p.1.}
\een

\medskip

{\em Proof.}
We expand
\ben
p_n(X_1^n,\delta)
\;=\;
\PR\{\rho_n(X_1^n,Y_1^n)\leq D\;\mbox{and}\;\Phatn\in B_\delta \}
        \Big/
    \PR\{\Phatn\in B_\delta \}
\een
and evaluate the exponential
behavior of the numerator and denominator
separately. First, by Sanov's theorem
\cite{csiszar:84},
\be
\lim_{n\to\infty}
\frac{1}{n}\log
    \PR\{\Phatn\in B_\delta \}
    = -R(\delta).
\label{eq:denominator}
\ee
We will also show that
\be
\lim_{n\to\infty}
\frac{1}{n}\log
\PR\{\rho_n(X_1^n,Y_1^n)\leq D\;\mbox{and}\;\Phatn\in B_\delta \}
    = -R(P,Q,D)
    \;\;\;\;\mbox{w.p.1,}
\label{eq:numerator}
\ee
and, recalling that
$R(P,Q,D)-R(\delta)=I(\delta)$,
this will complete the proof.

First note that, since
$$\PR\{\rho_n(X_1^n,Y_1^n)\leq D\;\mbox{and}\;\Phatn\in B_\delta \}
\leq Q^n(B(X_1^n,D)),$$
Theorem~B implies
\be
\limsup_{n\to\infty}
\frac{1}{n}\log
\PR\{\rho_n(X_1^n,Y_1^n)\leq D\;\mbox{and}\;\Phatn\in B_\delta \}
    \leq -R(P,Q,D)
    \;\;\;\;\mbox{w.p.1.}
\label{eq:numeratorUB}
\ee
For the corresponding lower bound we
employ the G\"{a}rtner-Ellis theorem,
much as in the proof of Theorem~3.
Let $x_1^\infty$ be some fixed
realization
of $\Xp$, and define a
sequence of random vectors $\{\zeta_n\}$
in $\RL^{k+1}$ by
$$\zeta_n\bydef \left(\rho_n(x_1^n,Y_1^n),\Phatn(F_1),\ldots,
    \Phatn(F_k)\right)
=\frac{1}{n}\sum_{i=1}^n\left(\rho(x_i,Y_i),
\IND_{F_1}(Y_i),
\ldots,
\IND_{F_k}(Y_i)
\right).$$
Let $\Gamma_n(\la)$ denote the log-moment generating
function of $\zeta_n$,
$$\Gamma_n(\la)\bydef\ln
    E_{Q^n}\left[
        \exp\left\{\la_0\rho_n(x_1^n,Y_1^n)
            +\sum_{i=1}^k\la_i\Phatn(F_i)\right\}
    \right],
    \;\;\;\;\;\;
    \la=(\la_0,\dots,\la_k)\in(-\infty,0]^{k+1}.$$
As before, the
ergodic theorem says that for $\BBP$-almost
every realization $x_1^\infty$,
\ben
    \lim_{n\to\infty} \frac{1}{n}\Gamma_n(n\la)
\;=\;
\Gamma(\la)
\;\bydef\;
E_P\left\{\ln E_Q\left[
                \exp\{\la_0\rho(X,Y)
            +\sum_{i=1}^k\la_i\IND_{F_i}(Y)\}
        \right]
    \right\},
\een
where, by Jensen's inequality,
the limiting moment generating
function $\Gamma(\la)$ satisfies
$-\infty<(\la_0\Dav+\sum_{i=1}^k\la_i)
\leq\Gamma(\la)\leq 0$. Once again,
a routine application
of the
dominated convergence
theorem verifies that $\Gamma(\la)$
is differentiable,
so the G\"{a}rtner-Ellis theorem
\cite[Theorem~2.3.6]{dembo-zeitouni:book}
yields that
\ben
\liminf_{n\to\infty}\frac{1}{n}\ln
\PR\{\rho_n(X_1^n,Y_1^n)\leq D\;\mbox{and}\;\Phatn\in B_\delta \}
\geq
-\inf_{z}\Gamma^*(z)
\;\;\;\;\mbox{w.p.1,}
\een
where
the infimum is taken over all
$z\in\RL^{k+1}$ with
$z_0\in[0,D)$ and
$|z_i-Q'(F_i)|<\delta$,
$i=1,\ldots,k$,
and $\Gamma^*$ denotes the convex dual of
$\Gamma$,
$$\Gamma^*(z)\bydef=\sup_{\la\in(-\infty,0]^{k+1}}
    [(z,\la)-\Gamma(\la)],
        \;\;\;\;z\in\RL^{k+1},$$
where $(z,\la)$ is the Euclidean inner
product in $\RL^{k+1}$, $(z,\la)=\sum_{i=0}^k z_i\la_i$.
Therefore, with probability one,
\be
\liminf_{n\to\infty}\frac{1}{n}\log
\PR\{\rho_n(X_1^n,Y_1^n)\leq D\;\mbox{and}\;\Phatn\in B_\delta \}
\geq
-(\log e)\Gamma^*(D,Q'(F_1),\ldots,Q'(F_k)),
\label{eq:LB2}
\ee
where we used the (easily verifiable)
fact that $\Gamma^*$ is continuous
in $z_0\in(\Dmin,\Dav)$.
Finally we claim that
\be
(\log e)\Gamma^*(D,Q'(F_1),\ldots,Q'(F_k))\leq R(P,Q,D).
\label{eq:claim}
\ee
Combining (\ref{eq:claim}) with (\ref{eq:LB2})
and with the upper bound (\ref{eq:numeratorUB})
proves (\ref{eq:numerator}) and the Lemma.

So it only remains to establish (\ref{eq:claim}).
Note that for any $x\in A$, any measurable function
$\phi:\Ahat\to\RL$ which is bounded above,
and any measure $W$ on $A\times\Ahat$,
\be
(\ln 2)H(W(\cdot|x)\|Q(\cdot))\geq \int\phi(y) W(dy|x)
-\ln E_Q[e^{\phi(Y)}].
\label{eq:DV}
\ee
[This can be proved in exactly the same way
as the corresponding statement in the proof
of \cite[Theorem~2]{dembo-kontoyiannis:wyner}.]
Take $W=W^*$ to be the achieving measure in the
definition of $R(P,Q,D)$, and
let $\phi(y)=\la_0\rho(x,y)+\sum_{i=1}^k\la_i\IND_{F_i}(y)$
for some $\la\in(-\infty,0]^{k+1}$. Applying
(\ref{eq:DV}) and integrating both sides
with respect to $P$ we get that
\ben
R(P,Q,D)
&\geq&
    (\log e)
    \left[\la_0 E_{W^*}[\rho(X,Y)]
    +\sum_{i=1}^k\la_i Q'(F_i)-\Gamma(\la)
    \right]\\
&\geq&
    (\log e)
    \left[(\la,(D,Q'(F_1),\ldots,Q'(F_k)))-
    \Gamma(\la)
    \right],
\een
and since this holds for any
$\la\in(-\infty,0]^{k+1}$ we have
established
(\ref{eq:claim}), as required.
\qed

\medskip

Finally we give a simple general result
on the asymptotic behavior of the entropy
of sequences of random variables.
Its proof is in Appendix~B.

\medskip

{\em Lemma~4}: Let $\xi_1,\xi_2,\ldots$ be
a sequence of random variables, and
$A_1,A_2,\ldots$be a sequence of events
with $\PR\{A_n\}\to 1$, as $n\to\infty.$
Assume that
$\xi_n\in\{1,2,3,\ldots,2^{n\beta}\}$, for all $n$
and some $\beta<\infty$. Then,
$$\lim_{n\to\infty}
\frac{1}{n}
\Big[H(\xi_n)-H(\xi_n|\IND_{A_n}=1)
\Big]=0.$$

\subsection{Proof of Theorem 1}
Let $\epsilon>0$ be arbitrary, and choose a $\delta>0$
and a finite partition ${\cal P}=\{F_1,\ldots,F_k\}$
of $\Ahat$ as in Lemma~2. With $B_\delta$ as in
(\ref{eq:bdelta}) and with $\Phatyk$ denoting the
empirical distribution of the $k$th codeword,
for $k=1,2,\ldots$ we define:
$$
J_k=\left\{
    \begin{array}{ll}
        \IND_{\{\Phatyk\in B_\delta\}}
            & \mbox{if $1\leq k \leq \lfloor 2^{nb}\rfloor$}, \\
        1
            & \mbox{if $k \geq \lfloor 2^{nb}\rfloor+1$}.
\end{array}
\right.
$$
Now we consider two sub-codebooks of $\Cp_n$,
\ben
\Cp_n^{(0)}
&\bydef&
    \left\{Y_1^n(k)\;:\;J_k=0,\;1\leq k\leq\lfloor 2^{nb}\rfloor\right\}\\
\Cp_n^{(1)}
&\bydef&
        \left\{Y_1^n(k)\;:\;J_k=1,\;1\leq k\leq\lfloor 2^{nb}\rfloor\right\}.
\een
Also,
for $j=0,1,$ let $N_n^{(j)}$ be the index of the first
codeword in $\Cp_n^{(j)}$ that matches $X_1^n$ with distortion
$D$ or less, and let $M_n^{(j)}$ be the index of the
position of $Y_1^n(N_n^{(j)})$ in $\Cp_n^{(j)}$.
If no match is found in $\Cp_n^{(j)}$, then let
$N_n^{(j)}=M_n^{(j)}=\lfloor 2^{nb}\rfloor+1$.  From
these definitions it immediately follows that,
given $\Cp_n$, the value of $N_n'$ and the values
of $(M_n^{(J_{N_n})}, J_{N_n})$ are in a one-to-one
correspondence.

To bound $E[\calL_n(X_1^n)]$
we begin by expanding
\ben
H(N_n'|\Cp_n)
&=&
    H(M_n^{(J_{N_n})}, J_{N_n} |\Cp_n)\\
&\leq&
    1+H(M_n^{(J_{N_n})}|J_{N_n})\\
&=&
    1
    + \PR\{J_{N_n}=0\}H(M_n^{(J_{N_n})}|J_{N_n}=0)
    + \PR\{J_{N_n}=1\}H(M_n^{(J_{N_n})}|J_{N_n}=1)\\
&\leq&
    1
        + \PR\{J_{N_n}=0\}\log(\lfloor 2^{nb}\rfloor+1)
        + H(M_n^{(1)}|J_{N_n}=1),
\een
therefore, in view of (\ref{eq:equivalence})
\be
\limsup_{n\to\infty} \frac{1}{n} E[\calL_n(X_1^n)]
&\leq&
    \limsup_{n\to\infty}
    \frac{1}{n} H(N'_n |\Cp_n)
    \nonumber\\
&\leq&
        \limsup_{n\to\infty}
    \left[\frac{1}{n}H(M_n^{(1)}|J_{N_n}=1)
        +\frac{1}{n}\log(2^{nb}+1)\PR\{\Qhatn\not\in B_\delta\}
    \right]
    \nonumber\\
&\eqa&
    \limsup_{n\to\infty}\frac{1}{n}H(M_n^{(1)}|J_{N_n}=1)
    \nonumber\\
&\eqb&
        \limsup_{n\to\infty}\frac{1}{n}H(M_n^{(1)}),
    \label{eq:bound1}
\ee
where~$(a)$ follows from Theorem~2, and~$(b)$ follows
from Lemma~4.  Now recall the
definition of the (conditional)
probability $p_n=p_n(X_1^n,\delta)$ from Lemma~3,
and for arbitrary $\Delta>0$ let $\expd$ denote
a quantized version of the exponent of $p_n$:
\ben
E_n^{(\Delta)}\bydef
\Delta\left\lceil-\frac{1}{n\Delta}\log p_n\right\rceil.
\een
Note that, given a source string $x_1^n$, the random
variable $M_n^{(1)}$ has a ``truncated Geometric''
distribution, which we denote by $\Gts(p_n)$;
formally, for a parameter $q\in(0,1)$,
$$
\PR\{\Gts(q)=k\}=
\left
\{ \begin{array}{ll}
   q(1-q)^{k-1} & \mbox{if $1\leq k\leq \lfloor 2^{nb}\rfloor$} \\
   (1-q)^{k-1}    & \mbox{if $k = 1+ \lfloor 2^{nb}\rfloor$} \\
   0        & \mbox{otherwise}.
\end{array}
\right.
$$
A useful bound on the
entropy of a mixture of $\Gts(q)$ distributions
is given in the following lemma;
its proof is given in Appendix~C.

\medskip

{\em Lemma~5}:
If the distribution
of the random variable $Z$ is a mixture of
$\Gts(q)$ distributions with $q\in[\alpha,\beta]$,
then $H(Z)\leq  \log(e/\alpha).$

\medskip

Now observe that, given $\expd$ is equal to some $e_n$,
the conditional distribution of
$M_n^{(1)}$ is a mixture of $\Gts(q)$ distributions, for
parameter values $q\geq 2^{-ne_n}$. Therefore, by Lemma~5,
$$H(M_n^{(1)}|\expd=e_n)\leq  n e_n + \log(e),$$
and hence, with probability one,
\ben \limsup_{n\to\infty}\frac{1}{n}H(M_n^{(1)}|\expd=e_n)
&\leq&
    \limsup_{n\to\infty}  \expd  \\
&\leq&
    \limsup_{n\to\infty}-\frac{1}{n}\log p_n(X_1^n,\delta)+\Delta\\
&\leqa&
    I(\delta)+\Delta\\
&\leqb&
    I_m(P\|Q',D)+\epsilon+\Delta,
\een
where~$(a)$ follows by Lemma~2 and~(b) by Lemma~3.
Since for all $n$ large enough
$(1/n)H(M_n^{(1)}|\expd=e_n)\leq (b+1/n)\leq (b+1)$ with
probability one,
we can apply Fatou's lemma to get,
\be
\limsup_{n\to\infty}\frac{1}{n}H(M_n^{(1)}|\expd)
&=&
    \limsup_{n\to\infty}
    E\Big\{ \frac{1}{n}H(M_n^{(1)}|\expd=e_n)\Big\}
    \nonumber\\
&\leq&
    E\Big\{
    \limsup_{n\to\infty}
    \frac{1}{n}H(M_n^{(1)}|\expd=e_n)\Big\}
    \nonumber\\
&\leq&
    I_m(P\|Q',D)+\epsilon+\Delta.
\label{eq:bound2}
\ee
Next we will show that
\be
\limsup_{n\to\infty}\frac{1}{n}H(M_n^{(1)})
\leq
\limsup_{n\to\infty}\frac{1}{n}H(M_n^{(1)}|\expd).
\label{eq:bound3}
\ee
Since $\epsilon>o$ and $\Delta>0$ were arbitrary,
combining this with (\ref{eq:bound1}) and (\ref{eq:bound2})
will complete the proof of the theorem.

Turning to the proof of (\ref{eq:bound3}),
we take $\epsilon'>0$ arbitrary, and define
\be
I_n=\IND_{\{
    \expd\leq I_m(P\|Q',D)+\Delta+\epsilon+\epsilon'
    \}},
\label{eq:3e}
\ee
and observe that, by Lemmas~2 and~3,
\be
\PR\{I_n=1\}\geq
\PR\{-\smalloneovern\log p_n(X_1^n,\delta)\leq I(\delta)+\epsilon'\}\to 1
\;\;\;\;\mbox{as}\;n\to\infty.
\label{eq:prob}
\ee
We can expand
\ben
H(M_n^{(1)}|\expd)
&\geq&
    H(M_n^{(1)}|I_n,\expd)\\
&\geq&
    \PR\{I_n=1\}\,H(M_n^{(1)}|I_n=1,\expd)\\
&=&
    \PR\{I_n=1\}\left[
    H(M_n^{(1)},\expd|I_n=1)-
    H(\expd|I_n=1)
    \right]\\
&\geq&
    \PR\{I_n=1\}\left[
        H(M_n^{(1)},\expd|I_n=1)-K
        \right]\\
&\geqa&
    H(M_n^{(1)}|I_n=1) + (\PR\{I_n=1\}-1)\log(\lceil 2^{nb}\rceil+1)
    +K\PR\{I_n=1\}
\een
where
$$K=\log\left(\frac{I_m(P\|Q',D)+\Delta+\epsilon+\epsilon'}{\Delta}
    +1 \right)+1$$
and where (a) follows since by
(\ref{eq:3e}) the number of values of
$\expd$ is at most $2^K$.
Therefore, from this and (\ref{eq:prob}) we have
$$\limsup_{n\to\infty}\frac{1}{n}H(M_n^{(1)}|I_n=1)
    \leq \limsup_{n\to\infty}\frac{1}{n}
    H(M_n^{(1)}|\expd).$$
Finally, (\ref{eq:prob}) allows us to apply
Lemma~4 and thus conclude that (\ref{eq:bound3})
holds, completing the proof of the theorem.
\qed


\section{Tighter Bounds for Sources with Memory}
\label{section:last}

In this section we discuss the following questions:
(1)~Under what conditions is the bound in Theorem~1 tight?
(2)~When it is not tight, what is the actual performance of
the entropy-coded scheme? {\em Only heuristic arguments and
proof outlines are given.}

To gain some intuition, we first consider the
extreme case of {\em lossless} compression of
a finite-alphabet, stationary ergodic source $\Xp$,
that is $D=0$ relative to Hamming distortion.
Let $Q$ be a codebook distribution on $\Ahat=A$
with $Q(a)>0$ for all $a\in A$, and let $\Cp_n$
be a {\em memoryless} random codebook with distribution
$Q$. Then all possible $n$-strings from $A^n$ will
appear infinitely often in $\Cp_n$, and the matching
codeword will always be identical to the source string.
Moreover, this also implies that $Q^*_{PQD}$,
the limiting first-order empirical distribution of the matching
codeword (Theorem~2), will simply
be the first-order marginal $P$ of the source.
It is therefore an immediate consequence of the AEP
(Asymptotic Equipartition Property \cite{cover:book})
that the asymptotic rate achieved by this scheme
will be exactly equal to the entropy rate $H(\Xp)$ of
the source $\Xp$. In this case it is easy to calculate
the bound given in Theorem~1 explicitly to get that,
at $D=0$,
$$I_m(P\|Q^*_{PQD},D)= I_m(P\|P,D) = H(X_1).$$

The above argument indicates that the bound in Theorem~1
will be tight {\em if and only if} the source $\Xp$ is
memoryless. Indeed, for finite-alphabet memoryless sources
this was shown to be the case in \cite{zamir:index}.

Now let us turn to general alphabet sources and positive
distortions $D > 0$.
For general stationary sources, it is well-known
that the rate-distortion function decreases
as the memory increases,
so it is natural to expect that the rate achieved by any ``good''
coding scheme will also take advantage of
such dependencies.

For the {\em naive coding scheme} Theorem~A
immediately shows that,
if the codebook distribution is memoryless,
then memory in the source does {\em not} affect the rate
achieved.
Formally, this observation is reflected
in the identity \cite{yang-kieffer:1},
\begin{equation}
\label{Rpqd_k}
R(P_k, Q^k, kD) = k R(P_1,Q,D),
\;\;\;\;\mbox{for all $k$.}
\end{equation}
In contrast, in the {\em entropy-coded} case we expect that
memory in the source {\em does} affect the rate.
For example, the above heuristic argument shows that for $D=0$
entropy-coding achieves the entropy-rate of the source,
and not just $H(X_1)$. But since the bound
$I_m(P_1\|Q^*_{P_1 Q D},D)$  in Theorem~1
only depends on the first-order marginal $P_1$,
memory in the source
does not affect it and therefore it cannot be tight in this
case. As we discuss next we can establish tighter bounds showing
that, in fact, entropy-coding the index {\em does} take advantage
of memory. This more desirable behavior is reflected in the
multi-dimensional behavior of the lower mutual information (LMI)
function: In contrast to (\ref{Rpqd_k}), whenever
$P_k\neq P^k$,
\ben
I_m(P_k \| Q^*_{P_k,Q^k,kD}, kD) < k I_m(P_1 \| Q^*_{P_1,Q,D},D),
\een
where $Q^*_{P_k,Q^k,kD}$ is the (unique) $k$-dimensional
distribution $\tilde{Q}_k$ that achieves $R(P_k, Q^k, kD)$
in (\ref{eq:achieve}).
Therefore, the LMI decreases due to memory
in the source even if the codebook is memoryless.

\medskip

{\em Example. Universal Gaussian Codebooks. }
To appreciate this decrease in LMI due to memory in the source,
consider a memoryless Gaussian codebook with large variance
$\tau^2$ and squared error distortion measure,
as in Section~\ref{s:gaussian}. For a real-valued
source $\Xp$ with zero mean and finite variance
$\sigma^2$, a straightforward
$k$-dimensional extension of Proposition~1 gives,
    \begin{equation}   \label{poliker}
    \lim_{\tau^2\to\infty} I_m(P_k \|Q^*_{P_k,Q^k,D},kD) =
    I(X_1^k; X_1^k + Z^{(k)}_D)
    \end{equation}
    where $X_1^k \sim P_k$, and
    $Z^{(k)}_D$ denotes an $\iid$
    $N(0,D)$ random vector independent of $X_1^k$.
Since $Z^{(k)}_D$ has a density we can write
$$I(X_1^k; X_1^k + Z^{(k)}_D) = h(X_1^k + Z^{(k)}_D) - k h(Z^{(1)}_D) .$$
If $X_1^k$ also has a density, then
for small $D$ this expression becomes
$h(X_1^k) - k h(Z^{(1)}_D) + o(1)$,
where $o(1) \to 0$ as $D \to 0$;
see \cite{linder-zamir:94}.
It follows that, for small $D$,
\[
I_m(P_1 \| Q^*_{P_1,Q,D},D) - \frac{1}{k} I_m(P_k \|Q^*_{P_k,Q^k,kD},kD) =
h(P_1) - \frac{1}{k} h(P^k) + o(1)
\]
where $\lim_{D \to 0} \lim_{\tau^2 \to \infty} o(1) = 0$.
That is, for small $D$ the LMI rate reduction relative to the
marginal case is asymptotically
\[
h(P_1) - \frac{1}{k} h(P^k) \to I(X_1; X_{-\infty}^0)
\]
as $k \to \infty$.
This is the information the past has about the present,
which for some sources can be very large.

\medskip

In general, the tighter bounds on the rate of the
entropy-coded scheme follow from natural $k$-dimensional
extensions of the results in Theorems~1 and 2.
As before, we restrict attention to memoryless
random codebooks with arbitrary distribution $Q$,
single-letter distortion measures, and stationary ergodic
sources. As in \cite{kontoyiannis:sphere:01}, the extension
to the case with memory follows by considering $k$-blocks
of super-symbols in the source and the codebook, but the
technicalities, although not particularly insightful,
are very involved. The reader will have probably been
convinced of this by seeing the proofs in the simpler
memoryless case. Under the same assumptions as in Theorems~1
and~2 (and perhaps under mild additional regularity conditions
on the source as in \cite{kontoyiannis:sphere:01}), we
obtain the following analogs.

\medskip

{\em Theorem~1-k.}: For any $k$ we have:
\ben
\limsup_{n\to\infty}\frac{1}{n} H(N'_n |\Cp_n)
\;\leq\;
\frac{1}{k} I_m(P_k \|Q^*_{P_k,Q^k,kD},kD)
\;\;\;\;\mbox{bits/symbol.}
\een

\medskip

{\em Theorem~2-k.}:
Let $\Qhatn^{(k)}$ denote the $k$th order empirical
distribution induced by the matching codeword
$Z_1^n\bydef Y_1^n(N_n)$ on $\Ahat$.
With probability one, for any $k$ we have:
$$\Qhat_n^{(k)} \Rightarrow Q^*_{P_k,Q^k,kD},\;\;\;\;\mbox{as $n\to\infty$.}$$

\medskip

Following standard arguments used in the analysis of
the rate-distortion function \cite{berger:book},
we can define the {\em LMI rate} as
\[
I_m(\BBP \| Q^*_{\BBP, Q^\infty, D}, D) \bydef
\inf_k \frac{1}{k} I_m(P_k \|Q^*_{P_k,Q^k,kD},kD) =
\lim_{k \to \infty} \frac{1}{k} I_m(P_k \|Q^*_{P_k,Q^k,kD},kD) .
\]
It follows that the best upper bound on the index entropy is
$I_m(\BBP \| Q^*_{\BBP, Q^\infty, D}, D)$, and we
conjecture that this bound is in fact tight, i.e.,
\[
\lim_{n\to\infty}\frac{1}{n} H(N'_n |\Cp_n) =
I_m(\BBP \| Q^*_{\BBP, Q^\infty, D}, D) .
\]


{\em Example. Universal Gaussian Codebooks. }
Returning to the special case considered in the
last example, if $Q\sim N(0,\tau^2)$,
then
as the codebook variance $\tau^2\to\infty$
the rate $I_m(\BBP \| Q^*_{\BBP, Q^\infty, D}, D)$
achieved by the entropy-coded scheme satisfies
\be
\lim_{\tau^2\to\infty} I_m(\BBP \| Q^*_{\BBP, Q^\infty, D}, D)
        =I(\Xp; \Xp +\Zp_D),
    \label{eq:mem}
\ee
where $\Zp_D$ is a white Gaussian process with variance $D$
and $I(\Xp; \Xp +\Zp_D)$ is the mutual information rate
between $\Xp$ and $\Zp_D$.
(A simple heuristic calculation indicating that (\ref{eq:mem})
holds is to divide (\ref{poliker}) by $k$ and take $k$ to infinity.)
Combining this with the fact
that $I(\Xp; \Xp +\Zp_D)\leq R(D)+1/2$,
\cite{ziv:1,feder-zamir:92},
where $R(D)$  is the rate-distortion function of the entire process
(not just the first-order rate-distortion function),
we get that, as $\tau^2\to\infty$, the rate achieved
by the entropy-coded scheme is no worse than
$R(D) + 1/2$ bits/symbol.


\section*{Acknowledgments}
We thank Zhiyi Chi for some interesting
technical conversations.

\section*{Appendix}

\appendix

\section{Proof Outline of Proposition~1}
Although the result of the proposition can be
obtained by little more than elementary calculus,
the calculations are rather lengthy so we only
give an outline of the proof here.

First observe that, in the notation of Section~\ref{prelims},
the log-moment generating $\LA(\la)$ can be evaluated
explicitly,
$$\LA(\la)=-\frac{1}{2}\ln(1-2\la\tau^2)
    +\frac{\la\sigma^2}{1-2\la\tau^2},$$
and (\ref{eq:condition}) can be solved to show that
the optimizing value of $\la=\la^*$ is
given by
$$\la=\frac{2D-\tau^2-\Delta}{4\Delta\tau^2},$$
where
$$\Delta=\sqrt{\tau^2+4\sigma^2D},$$
so that, as $\tau^2\to\infty$ we have
\be
\la
&=&
    -\frac{1}{2D}+\frac{1}{2\tau^2}+O(\tau^{-4})
    \label{eq:la1}\\
\la^2
&=&
    \frac{1}{4D^2}+O(\tau^{-4}).
    \label{eq:la2}
\ee From
the proof of \cite[Theorem~2]{dembo-kontoyiannis:wyner}
it follows that the joint distribution $W^*$ that achieves
the infimum in the definition of $R(P,Q,D)$ is given by
$$
  \frac{dW^*}{d(P\times Q)}(x,y)=\frac{e^{\la(x-y)^2}}
  {E_P\left[e^{\la(X-y)^2}\right]},
$$
and, as discussed in Section~\ref{prelims},
$Q'$ is the $Y$-marginal $W^*_Y$ of $W^*$.
Therefore, writing $\phi_a(y)$ for the $N(0,a)$ density,
the density $f_{Q'}(y)$ of $Q'$ with respect
to Lebesgue measure $m(dy)$ can be expressed as
$$f_{Q'}(y)
\;=\;
    \frac{dQ'}{dm}(y)
\;=\;
    E_P\left[
    \frac{e^{\la(X'-y)^2}}
    {E_P\left[e^{\la(X-y)^2}\right]}
    \right]\,\phi_{\tau^2}(y),$$
where $X$ and $X'$ denote two independent
random variables with the same distribution
$P$.  Evaluating the denominator explicitly
and rearranging terms, the above expression becomes
\be
f_{Q'}(y)
=
\sqrt{\frac{\frac{1}{2\tau^2}-\la}{\pi}}\;
E_P
\left[
  \exp
  \left\{
    -\left(
    y\sqrt{\frac{1}{2\tau^2}-\la}
    -X\sqrt{\frac{\la^2}{\frac{1}{2\tau^2}-\la}}
     \right)^2
  \right\}
\right],
\label{eq:dom1}
\ee
and recalling (\ref{eq:la1}) and (\ref{eq:la2}),
we can let $\tau^2\to\infty$ to get that
\be
\frac{dW^*}{d(P\times Q)}(x,y)\,\frac{dQ}{dm}(y)
&=&
\sqrt{\frac{\frac{1}{2\tau^2}-\la}{\pi}}\;
  \exp
  \left\{
    -\left(
        y\sqrt{\frac{1}{2\tau^2}-\la}
        -x\sqrt{\frac{\la^2}{\frac{1}{2\tau^2}-\la}}
     \right)^2
  \right\}
    \nonumber\\
&\to&
    \phi_D(y-x)
\;\;\;\;
        \mbox{as $\tau^2\to\infty.$}
    \label{eq:dom2}
\ee
Invoking the
dominated convergence theorem we can
conclude that
\be
f_{Q'}(y)\to
E_P[\phi_D(y-X)]\;\;\;\;
        \mbox{as $\tau^2\to\infty,$}
\label{eq:propa}
\ee
as claimed. This proves~$(a)$.

For part~$(b)$ note that, from
(\ref{eq:achieve}) and the above discussion
it follows that
\ben
I_m(P\|Q,D)
&=&
    R(P,Q',D)-H(Q'\|Q)\\
&=&
    H(W^*\|P\times Q')-H(Q'\|Q)\\
&=&
        H(W^*\|P\times Q)\\
&=&
    \int
    \left[\frac{dW^*}{d(P\times Q)}(x,y)\frac{dQ}{dm}(y)
    \ln\left\{
    \frac{dW^*}{d(P\times Q)}(x,y)\frac{dQ}{dm}(y)
    \left(\frac{dQ'}{dm}(y)\right)^{-1}
    \right\}\right]
    dP(x)dm(y).
\een
Using the expressions for the densities
in (\ref{eq:dom1}) and (\ref{eq:dom2}),
recalling that $dQ/dm(y)=\phi_{\tau^2}(y)$,
and applying the convergence bounds
in (\ref{eq:la1}), (\ref{eq:la2}) and (\ref{eq:propa}),
it is straightforward to show that
the last integrand in $[\cdots]$
above converges to
$$\phi_D(y-x)\ln\left(\frac{\phi_D(y-x)}{E_P[\phi_D(y-X)]}\right).$$
Writing $V_D$ for the joint distribution of
the random variables $(X,X+Z_D)$ as in the statement
of the proposition, and $Q_D$ for the distribution of
$(X+Z_D)$, the above expression can be rewritten as
$$
\frac{dV_D}{d(P\times m)}(x,y)
\ln\left\{\frac{dV_D}{d(P\times Q_D)}(x,y)\right\}.
$$
Finally, using (\ref{eq:la1}) and (\ref{eq:la2})
to justify the use of the dominated convergence
theorem we get that also the integrals converge,
i.e., as $\tau^2\to\infty$,
\ben
I_m(P\|Q,D)
&=&
    \int
        \left[\frac{dW^*}{d(P\times Q)}(x,y)\frac{dQ}{dm}(y)
        \ln\left\{
        \frac{dW^*}{d(P\times Q)}(x,y)\frac{dQ}{dm}(y)
        \left(\frac{dQ'}{dm}(y)\right)^{-1}
        \right\}\right]
        dP(x)dm(y)\\
&\to&
    \int
    \frac{dV_D}{d(P\times m)}(x,y)
    \ln\left\{\frac{dV_D}{d(P\times Q_D)}(x,y)\right\}
    dP(x)dm(y)\\
&=&
    H(V_D\|P\times Q_D)\\
&=&
    I(X;X+Z_D),
\een
proving~$(b)$.

Finally part~(c) follows from the well-known
fact \cite{ziv:1,feder-zamir:92,zamirWZ}
that the rate-distortion function $R(D)$
of a real-valued memoryless source
(with respect to squared error distortion)
is bounded below
by $I(X;X+Z_D)-1/2$.
\qed

\section{Proof of Lemma~4}
First observe that
$$H(\xi_n)\leq H(\xi_n,\IND_{A_n})
= H(\xi_n)+H(\IND_{A_n}|\xi_n)\leq H(\xi_n)+1,$$
so that
\be
\lim_{n\to\infty}
\frac{1}{n}
\Big[H(\xi_n)-H(\xi_n,\IND_{A_n})
\Big]=0.
\label{eq:lem4a}
\ee
Also we can expand
\ben
\frac{1}{n}H(\xi_n,\IND_{A_n})
&=&
    \frac{1}{n}H(\IND_{A_n})
    +\frac{1}{n}H(\xi_n|\IND_{A_n}=1)\PR\{A_n\}
    +\frac{1}{n}H(\xi_n|\IND_{A_n}=0)(1-\PR\{A_n\})\\
&=&
    O(\smalloneovern)
    +\frac{1}{n}H(\xi_n|\IND_{A_n}=1)
    +(1-\PR\{A_n\})\frac{1}{n}\Big[
    H(\xi_n|\IND_{A_n}=0)-H(\xi_n|\IND_{A_n}=1)
    \Big]\\
&=&
    \frac{1}{n}H(\xi_n|\IND_{A_n}=1)
    +O(\smalloneovern)
    +(1-\PR\{A_n\})\frac{1}{n}\log(2^{n\beta}),
\een
i.e.,
\be
\lim_{n\to\infty}
\frac{1}{n}
\Big[H(\xi_n,\IND_{A_n})-H(\xi_n|\IND_{A_n}=1)
\Big]=0.
\label{eq:lem4b}
\ee
Combining (\ref{eq:lem4a}) with (\ref{eq:lem4b})
proves the lemma.
\qed

\section{Proof of Lemma~5}
It is well-known that the Geometric (non-truncated) distribution
has the largest entropy among all nonnegative variables with a
given mean.
Now, it is easy to verify that if $Z_q$ is a Geom($q$)
random variable [i.e., if $Z_q=k$ with probability
$q(1-q)^{k-1}$, for $k=1,2,\ldots$],
then $E[Z_q] = 1/q$,
and
$$ H(Z_q) = \log(1/q) - \frac{1-q}{q} \log(1-q) \leq \log(e/q) . $$
Thus, since the mean of a mixture of truncated Geometric distributions
is smaller than or equal to the mean of the Geometric distribution
with the smallest parameter $q$ (in our case, $\alpha$),
we obtain that
$H(Z) \leq H(Z_\alpha) \leq \log(e/\alpha)$.
\qed

\section{Proof of Proposition~2}
We first introduce some convenient notation.
Let $W^*_s$  denote
the joint distribution minimizing  $H(W \| P \times Q_s)$ in
(\ref{eq:Rdefn})
(i.e., achieving $R(P,Q_s,D)$),
and let $Q^*_s$ denote its induced $Y$-marginal.
In our previous notation,
$Q^*_s = Q_{P,Q_s,D}$
and $I(W^*_s) = I_m(P||Q_{P,Q_s,D}, D)$,
where $I(W)$ is the mutual information associated with a joint
distribution $W$.
Let $\Wadd$ denote the joint input-output distribution associated with
the additive noise channel
$Y = X + Z_D$
in (\ref{additive-noise-channel}).
In this notation, part~(ii) of the proposition
amounts to
\[
I_m(P\|Q_{P,Q_s,D}, D) = I(W^*_s)   \rightarrow  I(\Wadd)
\]
as
$s \rightarrow \infty$.

Clearly $\Wadd$ is in the set of admissible
distributions $W$ in (\ref{eq:Rdefn}).
Moreover,
by Lemma~TIGHT we know that
$H(\Wadd \| P \times Q_s) - R(P,Q_s,D)  \rightarrow 0$
as
$s \rightarrow \infty$.
That is, $\Wadd$ asymptotically achieves the minimum of
$H(W||P\times Q_s)$.
Since, by (\ref{eq:Rdefn}) and the Pythagorean theorem for divergence
\cite{cover:book}
for any admissible $W$
\[
H(W \| P \times Q_s) \geq R(P,Q_s,D)  +  H(W^*_s \| W) ,
\]
we conclude that
\begin{equation}  \label{*}
H(W^*_s \| \Wadd) \rightarrow 0.
\end{equation}
Since relative entropy dominates $L_1$ distance,
this implies that the density of $W^*_s$ converges to that of
$\Wadd$, {\em a fortiori} proving part (i).

Part (i) and the semi-continuity of the divergence
\cite{pinsker:book}
imply that
\begin{equation}
\label{**}
\liminf_{s\to 0} I(W^*_s) \geq I(\Wadd) .
\end{equation}
On the other hand, it follows from (\ref{*}) and the chain rule
for relative entropy, \cite{cover:book}, that the
conditional relative entropy
$H(W^*_s \| \Wadd | P)  \rightarrow 0$ as $s\to\infty$.
Alternatively, if we expand $H(W^*_s \| \Wadd | P)$ in
terms of differential entropy, this becomes
%
%
%
$$\lim_{s\to 0} \,[h(Z_D) - h(Y^*_s|X)]=0$$
where  $(X,Y^*_s)$ are jointly distributed as $W^*_s$,
$Z_D$ is a maximum entropy random variable independent of $X$,
and $E{\rho(Y^*_s - X)} = D$
(equality here is due to the strict monotonicity
of $\hmax(D)$ as a function of $D$).
Therefore,
\begin{equation} \label{****}
h(Y^*_s|X)  \rightarrow  h(Z_D) = \hmax(D)   .
\end{equation}
We can also conclude from (\ref{*}) that
the relative entropy
between the outputs vanishes
\[
\lim_{s\to 0}H(Q^*_s \| Q_Y) = 0,\\
\]
where $Q_Y$ denotes the distribution of $Y=X+Z_D$.
Again by the semi-continuity of the divergence this implies
that
\begin{equation}
\label{*****}
\limsup_{s\to 0} h(Y^*_s) \leq h(X+Z_D);
\end{equation}
see \cite{linder-zamir:94}.
Combining (\ref{****}) and (\ref{*****}) we thus have
\begin{eqnarray}
I(W^*_s) &=& I(X; Y^*_s)  \\
             &=& h(Y^*_s) - h(Y^*_s|X)  \\
             &\leq& h(X+Z_D) - h(Z_D) + o(1)  \\
             &=& I(X; X + Z_D)  +o(1)
\end{eqnarray}
where $o(1) \rightarrow 0$ as $s \rightarrow \infty$.
This, together with (\ref{**}), proves part (ii).

Finally, part (iii) follows from \cite{zamirWZ}.
\qed



\begin{thebibliography}{10}

\bibitem{berger:book}
T.~Berger.
\newblock {\em Rate Distortion Theory: A Mathematical Basis for Data
  Compression}.
\newblock Prentice-Hall Inc., Englewood Cliffs, NJ, 1971.

\bibitem{billingsley:cpm}
P.~Billingsley.
\newblock {\em Convergence of Probability Measures}.
\newblock John Wiley \& Sons Inc., New York, second edition, 1999.

\bibitem{bucklew:84}
J. A. Bucklew.
\newblock Two results on the asymptotic performance of quantizers.
\newblock {\em IEEE Trans. Inform. Theory}, 30:341--348, March 1984.

\bibitem{chou-l-gray}
P.A. Chou, T.~Lookabaugh, and R.M. Gray.
\newblock Entropy constrained vector quantization.
\newblock {\em IEEE Trans. Acoustics, Speech and Signal Processing}, 37:31--42,
  1989.

\bibitem{cover:book}
T.M. Cover and J.A. Thomas.
\newblock {\em Elements of Information Theory}.
\newblock J. Wiley, New York, 1991.

\bibitem{csiszar:75}
I.~Csisz{\'{a}}r.
\newblock ${I}$-divergence geometry of probability distributions and
  minimization problems.
\newblock {\em Ann. Probab.}, 3:146--158, 1975.

\bibitem{csiszar:84}
I.~Csisz{\'a}r.
\newblock Sanov property, generalized ${I}$-projection and a conditional limit
  theorem.
\newblock {\em Ann. Probab.}, 12(3):768--793, 1984.

\bibitem{dembo-kontoyiannis:wyner}
A.~Dembo and I.~Kontoyiannis.
\newblock Source coding, large deviations, and approximate pattern matching.
\newblock {\em IEEE Trans. Inform. Theory}, 48:1590--1615, June 2002.

\bibitem{dembo-zeitouni:book}
A.~Dembo and O.~Zeitouni.
\newblock {\em Large Deviations Techniques And Applications}.
\newblock Springer-Verlag, New York, second edition, 1998.

\bibitem{elias}
P.~Elias.
\newblock Universal codeword sets and representations of the integers.
\newblock {\em IEEE Trans. Inform. Theory}, 21:194--203, 1975.

\bibitem{gersho:79}
A.~Gersho.
\newblock Asymptotically optimal block quantization.
\newblock {\em IEEE Trans. Inform. Theory}, 25:373--380, 1979.

\bibitem{gish-pierce:68}
H.~Gish and N.J. Pierce.
\newblock Asymptotically efficient quantization.
\newblock {\em IEEE Trans. Inform. Theory}, 14:676--683, 1968.

\bibitem{gray-linder:03}
R.M.~Gray and T. Linder.
\newblock Mismatch in high-rate entropy-constrained vector quantization.
\newblock {\em IEEE Trans. Inform. Theory}, 49:1204--1218, 2003.

\bibitem{gutman:87}
M.~Gutman.
\newblock On universal quantization with various distortion measures.
\newblock {\em IEEE Trans. Inform. Theory}, 33: Jan. 1987.

\bibitem{kanlis:phd}
A.~Kanlis.
\newblock {\em Compression and Transmission of Information at Multiple
  Resolutions}.
\newblock PhD thesis, Dept. of Electrical and Computer Engineering, University
  of Maryland at College Park, 1998.

\bibitem{kanlis-narayan-rimoldi}
A.~Kanlis, P.~Narayan, and B.~Rimoldi.
\newblock On three topics for a course in information theory.
\newblock In {\em Statistical Methods in Imaging, Medicine, Optics, and
  Communication, Edt. J.A. O'Sullivan}. Springer Verlag, 2001.

\bibitem{kieffer:91}
J.C. Kieffer.
\newblock Sample converses in source coding theory.
\newblock {\em IEEE Trans. Inform. Theory}, 37(2):263--268, 1991.

\bibitem{kontoyiannis:sphere:01}
I.~Kontoyiannis.
\newblock Sphere-covering, measure concentration, and source coding.
\newblock {\em IEEE Trans. Inform. Theory}, 47:1544--1552, May 2001.

\bibitem{konto-zhang:02}
I.~Kontoyiannis and J.~Zhang.
\newblock Arbitrary source models and {B}ayesian codebooks in rate-distortion
  theory.
\newblock {\em IEEE Trans. Inform. Theory}, 48:2276--2290, 2002.

\bibitem{lapidoth:97}
A.~Lapidoth.
\newblock On the role of mismatch in rate distortion theory.
\newblock {\em IEEE Trans. Inform. Theory}, 43(1):38--47, 1997.

\bibitem{linder-zamir:94}
T.~Linder and R.~Zamir.
\newblock On the asymptotic tightness of the {S}hannon lower bound.
\newblock {\em IEEE Trans. Inform. Theory}, 40(6):2026--2031, 1994.

\bibitem{lookabaugh-gray:89}
T.~Lookabaugh and R.M. Gray.
\newblock High resolution quantization theory and the vector quantizer
  advantage.
\newblock {\em IEEE Trans. Inform. Theory}, 35:1020--1033, 1989.

\bibitem{neuhoff:1}
D.L. Neuhoff.
\newblock Source coding strategies: {S}imple quantizers vs. simple noiseless
  codes.
\newblock {\em Proceedings 1986 Conf. on Information Sciences and Systems},
  1:267--271, 1986.


\bibitem{pinkston:67}
J.T. Pinkston.
\newblock {\em Encoding Independent Sample Information Sources}.
\newblock PhD Thesis, MIT, 1967.


\bibitem{pinsker:book}
M.S. Pinsker.
\newblock {\em Information and Information Stability of Random Variables and
  Processes}.
\newblock Holden-Day, San Francisco, 1964.


\bibitem{sakrison:69}
D.J. Sakrison.
\newblock The rate distortion function for a class of sources.
\newblock {\em Information and Control}, 15:165--195, 1969.

\bibitem{sakrison:70}
D.J. Sakrison.
\newblock The rate of a class of random processes.
\newblock {\em IEEE Trans. Inform. Theory}, 16:10--16, 1970.

\bibitem{steinberg-gutman}
Y.~Steinberg and M.~Gutman.
\newblock An algorithm for source coding subject to a fidelity criterion, based
  upon string matching.
\newblock {\em IEEE Trans. Inform. Theory}, 39(3):877--886, 1993.

\bibitem{yang-kieffer:1}
E.-h. Yang and J.C. Kieffer.
\newblock On the performance of data compression algorithms based upon string
  matching.
\newblock {\em IEEE Trans. Inform. Theory}, 44(1):47--65, 1998.

\bibitem{yang-kieffer:96}
E.-h. Yang and J.C. Kieffer.
\newblock Simple universal lossy data compression schemes derived
from the Lempel-Ziv algorithm.
\newblock {\em IEEE Trans. Inform. Theory}, 42:239--245, Jan. 1996.

\bibitem{zamir:index}
R.~Zamir.
\newblock The index entropy of a mismatched codebook.
\newblock {\em IEEE Trans. Inform. Theory}, 48(2):523--528, 2002.

\bibitem{feder-zamir:92}
R.~Zamir and M.~Feder.
\newblock On universal quantization by randomized uniform/lattice quantizers.
\newblock {\em IEEE Trans. Inform. Theory}, 38:428--436, 1992.

\bibitem{feder-zamir:96}
R.~Zamir and M.~Feder.
\newblock Information rates for pre/post filtered dithered  quantizers.
\newblock {\em IEEE Trans. Inform. Theory}, 42:1340--1353, 1996.

\bibitem{zamir-rose:01}
R.~Zamir and K.~Rose.
\newblock Natural type selection in adaptive lossy compression.
\newblock {\em IEEE Trans. Inform. Theory}, 47(1):99--111, 2001.

\bibitem{zamirWZ}
R.~Zamir.
\newblock The rate loss in the Wyner-Ziv problem.
\newblock {\em IEEE Trans. Inform. Theory}, 42:2073-2084, Nov. 1996.

\bibitem{zhang-wei:1}
Z.~Zhang and V.K. Wei.
\newblock An on-line universal lossy data compression algorithm by continuous
  codebook refinement -- {P}art~{I}: {B}asic results.
\newblock {\em IEEE Trans. Inform. Theory}, 42(3):803--821, 1996.

\bibitem{ziv:1}
J.~Ziv.
\newblock On universal quantization.
\newblock {\em IEEE Trans. Inform. Theory}, 31(3):344--347, 1985.

\end{thebibliography}

\end{document}